\documentclass[aps,twocolumn,floats,superscriptaddress,prd,nofootinbib]{revtex4-1}
\usepackage{graphicx, epsfig, bm, amsmath}

\usepackage{color}
\usepackage{hyperref}
\usepackage{ wasysym }
\usepackage{amssymb}

\begin{document}


\def\nn{\nonumber}
\def\({\left(}
\def\){\right)}
\def\[{\left[}
\def\]{\right]}
\def\<{\left\langle}
\def\>{\right\rangle}
\newcommand{\lcdm}{$\Lambda$CDM}

\newcommand{\gpr}{G^{\prime}}

\newcommand{\fnl}{f_{\rm NL}}
\newcommand{\curv}{{\cal R}}

\definecolor{darkgreen}{cmyk}{0.85,0.2,1.00,0.2}
\newcommand{\peter}[1]{\textcolor{red}{[{\bf PA}: #1]}}
\newcommand{\vin}[1]{\textcolor{darkgreen}{[{\bf VM}: #1]}}
\newcommand{\wh}[1]{{#1}}

\newcommand{\action}{\mathcal{S}}

\newcommand{\WP}{W}
\newcommand{\XP}{X}
\newcommand{\B}{B^{\rm Bulk}}
\newcommand{\gB}{g_B}
\newcommand{\R}{\mathcal{R}}
\newcommand{\dotR}{\dot{\mathcal{R}}}
\newcommand{\ddotR}{\ddot{\mathcal{R}}}
\newcommand{\ep}{\epsilon_H}
\newcommand{\dotep}{\dot{\epsilon}_H}
\newcommand{\et}{\eta_H}
\newcommand{\dotet}{\dot{\eta}_H}
\newcommand{\cs}{c_s}
\newcommand{\esq}{\left(}  
\newcommand{\dir}{\right)} 
\newcommand{\ord}{\mathcal{O}}
\newcommand{\hR}{\hat{\mathcal{R}}}
\newcommand{\dothR}{\dot{\hat{\mathcal{R}}}}
\newcommand{\bpi}{\bm{\pi}}
\newcommand{\bp}{{\bf p}}
\newcommand{\csss}{c_{s\star}^2}

\newcommand{\aap}{Astron. Astrophys.}


\pagestyle{plain}

\title{Bispectrum  in Single-Field Inflation Beyond Slow-Roll}

\author{Peter Adshead}
\affiliation{Kavli Institute for Cosmological Physics,  Enrico Fermi Institute, University of Chicago, Chicago, IL 60637}

\author{ Wayne Hu}
\affiliation{Kavli Institute for Cosmological Physics,  Enrico Fermi Institute, University of Chicago, Chicago, IL 60637}
\affiliation{Department of Astronomy \& Astrophysics, University of Chicago, Chicago IL 60637}

\author{  Vin\'icius Miranda }
\affiliation{Department of Astronomy \& Astrophysics, University of Chicago, Chicago IL 60637}
\affiliation{The Capes Foundation, Ministry of Education of Brazil, Bras\'ilia DF 70359-970, Brazil}

\begin{abstract}

We develop an integral form for the bispectrum in general single-field inflation whose domain of validity includes models of inflation where the background evolution is not constrained to be slowly varying everywhere. Our integral form preserves the squeezed-limit consistency relation, allows for fast evaluation of the bispectrum for all triangle configurations expediting the efficient comparison of slow-roll violating models with data, and provides complete and compact slow-roll expressions correct to first order in slow-roll parameters. Motivated by the recent  \emph{Planck} results, we consider as an example a sharp step in the warped-brane tension of DBI inflation and provide analytic solutions for the peak of the resulting bispectrum.   For the
step in the warp that reproduces the oscillations in the power spectrum favored by the \emph{Planck} data, the corresponding equilateral bispectrum is both
extremely large and highly scale dependent. The bispectrum serves as a means of distinguishing such a model from alternative scenarios that generate otherwise indistinguishable power spectra, such as a step in the potential in canonical single-field inflation.

\end{abstract}

\maketitle

\section{Introduction}
\label{sec:intro}

In this paper we develop a technique for calculating the bispectrum of primordial 
fluctuations in general theories of single-field inflation beyond the usual slow-roll assumption where all of the slow-roll parameters are considered to be small and constant.   
This technique is based on the 
 generalized slow-roll (GSR) approach \cite{Stewart:2001cd, Choe:2004zg, Kadota:2005hv, Dvorkin:2009ne, Hu:2011vr, Miranda:2012rm} which has been successfully applied to the bispectrum in canonical single field inflation \cite{Adshead:2011bw, Adshead:2011jq, Adshead:2012xz}.  Here we extend this treatment to also consider terms in the effective field
 theory of inflation associated with time variation of the sound speed or equivalently an inflaton Lagrangian that is a general function of the field and its kinetic term.  Non-Gaussianity in  such models can be doubly enhanced due to a low and variable sound speed.
 
 This type of enhanced non-Gaussianity can be realized in several different ways.
 In Dirac-Born-Infeld (DBI) inflation \cite{Silverstein:2003hf, Alishahiha:2004eh}, variations in the sound speed are associated  with features in the warped brane tension \cite{Hailu:2006uj,Bean:2008na}.   
 For example a step in the tension produces a step in the sound speed.
Step features in the sound speed can be considered in a more general context such as the effective field theory of inflation \cite{Park:2012rh}. 
  
 Furthermore, transient variation in the sound speed has been shown to arise naturally in effective field theories of inflation from scenarios within multi-field inflation where heavy fields are integrated out \cite{Achucarro:2012sm, Achucarro:2012yr} (see also \cite{ Gao:2012uq}). In these scenarios, turning trajectories in field space result in variations in the speed of sound of the fluctuations. 
 
 In both of these cases, if the sound speed variations are sharp enough they generate
 characteristic oscillations in   the curvature power spectrum \cite{Miranda:2012rm,Achucarro:2010da} and bispectrum \cite{Bean:2008na,Achucarro:2012fd}.  They represent specific
 cases of the general phenomenon that slow-roll
violations in single-field inflation lead to features in the spectra of curvature fluctuations  \cite{Adams:2001vc, Wang:1999vf, Chen:2006xjb, Chen:2008wn}.  Starobinsky first noted that a sharp change in the slope of the inflaton potential lead to oscillatory features in the power spectrum \cite{Starobinsky:1992ts} and it has been recently shown that these models have large scale-dependent bispectra \cite{Martin:2011sn, Arroja:2011yu, Arroja:2012ae}. Violation of slow-roll via a rapidly varying sinusoidal component of the potential leading to resonance effects in the correlation functions was first noted by \cite{Chen:2006xjb, Chen:2008wn} before it was found to arise naturally in axion-monodromy inflation \cite{Flauger:2009ab,  Hannestad:2009yx, Flauger:2010ja} (see also \cite{Behbahani:2011it,Behbahani:2012be}).  Further work on slow-roll violating models includes \cite{Joy:2007na,Hotchkiss:2009pj, Nakashima:2010sa,  Leblond:2010yq,Chen:2010bka, Chen:2011zf, Chen:2011tu, Hazra:2010ve, Hazra:2012yn}.

On the observational side,
oscillations in the curvature power spectrum due to transient violations of slow-roll  were invoked to explain broad glitches  in the  \emph{WMAP} cosmic microwave background (CMB) angular power
 spectrum  \cite{Peiris:2003ff, Covi:2006ci, Hamann:2007pa}.
 Intriguingly   there is also a slight preference in the \emph{WMAP} data for high frequency oscillations in the power spectrum  near the first peak \cite{Adshead:2011jq}. 
 {In the  \emph{Planck} data, this preference persists {out to higher multipoles} but at a somewhat suppressed amplitude
 and at a harmonic of approximately twice the period $\Delta \ell \approx 10$ \cite{Ade:2013rta}.}
  The high frequency oscillations in the power spectrum
 take the same form regardless of whether they came from a sharp step in the
 potential or sound speed and hence it does not distinguish between these possibilities \cite{Miranda:2012rm}.

 Different explanations for features in the power spectrum should be distinguishable in the
 angular bispectrum of the CMB.  
The techniques developed here allow efficient computation of the curvature bispectrum in all of these cases, from weak violations of the slow-roll approximation to nearly order unity violations.
CMB angular bispectrum constraints on oscillating  curvature bispectrum shapes were first considered by Ref.\ \cite{Fergusson:2010dm} using a modefunction expansion method \cite{Fergusson:2009nv} on  \emph{WMAP} data. However, the constraints presented there correspond to much lower frequency oscillations, for technical reasons.  A similar analysis was performed on the  \emph{Planck} data  \cite{Ade:2013tta}, where again only periods in the angular bispectrum of $\Delta \ell > 140$  were considered and no significant evidence for features was found. Neither of these analyses have explored the high frequency region of parameter space where a large non-Gaussian counterpart of the best fit power spectrum feature  would be expected.   {The computation of the curvature bispectrum template for matching power spectrum signatures represents the first 
step for the analysis of high frequency features in the CMB bispectrum.}

This paper is organized as follows. In \S \ref{sec:bispec}, we review the in-in formalism and cast the cubic order action in a form amenable to bispectrum evaluation. In \S \ref{sec:GSR} we employ the generalized slow-roll formalism to derive integral expressions for the bispectrum that arises in general single-field inflation. In \S \ref{sec:DBI} we consider the specific example of warp features in DBI inflation to illustrate our technique before concluding in \S \ref{sec:conclusions}.  In Appendix \ref{app:additionalterms} we complete the bispectrum description with operators that 
do not appear in DBI inflation as well as treat the remaining slow-roll suppressed contributions.   In Appendix \ref{app:sr} we use our formalism to derive a compact, complete
expression for the slow-roll bispectrum to first order in slow-roll parameters.
In Appendix \ref{app:parameters}, we give computational details for the DBI example.
We take units where $M_{\rm pl}= 1/\sqrt{8\pi G}=1$ throughout.

\section{Bispectrum}
\label{sec:bispec}

In this section we consider the bispectrum in a general single field model of inflation. These models are characterized
by a nearly time-translation invariant expansion history through the Hubble parameter
$H(t)$ and its associated slow-roll parameters
\begin{eqnarray}
\epsilon_H &\equiv & - \frac{1}{H}\frac{ d\ln H}{dt}, \nonumber\\
 \eta_H &\equiv & \epsilon_H - \frac{1}{2H } \frac{d\ln\epsilon_H}{dt},
\end{eqnarray}
combined with the sound speed of inflaton fluctuations $c_s(t)$ and  its associated slow-roll parameter
\begin{eqnarray}
 		\sigma_1 & \equiv & \frac{1}{H} \frac{d\ln c_s}{dt} .
\end{eqnarray}
Beyond the slow-roll approximation, the parameters $\epsilon_H$, $\eta_H$ and $\sigma_1$
are allowed to vary with time as long as inflation itself continues without interruption
$\epsilon_H \ll 1$.    We establish our bispectrum formalism in terms of these general functions
that are specified by the model.

We begin in \S \ref{sec:inin} with a brief review of the `in-in' formalism for the calculation of
correlation functions and in \S \ref{sec:cubicaction} we show how the cubic action can be written in a form that allows a
straightforward calculation of all bispectrum configurations for models where the slow-roll parameters
are allowed to evolve.

\subsection{In-In Formalism}\label{sec:inin}

We work in the ``in-in'' formalism which expresses the $N$-point correlation function, or more generally,  the expectation value of a product of field operators $O(t)$ as \cite{Maldacena:2002vr,Weinberg:2005vy}
\begin{align}\label{eqn:inin}
\langle O(t_*\rangle = \langle U^{\dagger}(t_*, t_{0})O(t)U(t_*, t_{0})\rangle,
\end{align}
where $U(t_*, t_{0})$ is the time evolution operator in the interaction picture,
\begin{align}
U(t_*, t_{0}) = T\exp\left( - i\int_{t_0}^{t_*}H_{I}(t) dt\right),
\end{align}
and $H_I$ is the interaction Hamiltonian.  We take the initial time $t_{0}$ to be in the asymptotic past, $t_{0} = -\infty(1+i\varepsilon)$, where the $i\varepsilon$ prescription projects out the Bunch-Davies state initially.  We take the final time $t_*$ to be an arbitrary epoch during inflation after all of the relevant modes have exited the horizon.

 For the curvature bispectrum $B_\curv$, we wish to compute the  correlator 
\begin{equation}
\langle \hat{\curv}_{{\bf k}_1} \hat{\curv}_{{\bf k}_2} \hat{\curv}_{{\bf k}_3}\rangle
= (2\pi)^3 \delta({\bf k}_1+ {\bf k}_2+{\bf k}_3) B_{\curv}(k_1,k_2,k_3),
\end{equation}
where $\hat \curv_{\bf k}$ is the Fourier transform of the curvature field in comoving gauge
\begin{align}
\hat \curv({\bf x}, t) = \int \frac{d^{3}k}{(2\pi)^3}e^{i {\bf k}\cdot{\bf x}} \hat\curv_{{\bf k}}( t),
\end{align}
and the hat ($\hat{\curv}$) denotes an operator to differentiate it from the mode function ($\curv$).
In the interaction picture, the curvature fluctuations $\curv$ evolve according to their
quadratic action
\begin{eqnarray}
\action_2 &\equiv & \int dt d^3 x \mathcal{L}_2 
\nonumber\\ &=& \int dt d^3 x \frac{a^3 \epsilon_H}{c_{s}^2}\left[ \dot\curv^2 - \frac{c_s^2}{a^2}\(\partial \curv\)^2\right],
\label{eqn:L2}
\end{eqnarray}
yielding the equation of motion for $\curv$
\begin{equation}
\frac{d}{dt}\(\frac{a^3 \epsilon_H}{c_s^2}\dot\curv\) - a \epsilon_H\partial^{2}\curv=0.
\label{eqn:EOM}
\end{equation}

The tree-level bispectrum is then given by expanding Eq.~(\ref{eqn:inin}) [with $O(t_*) = \hat \curv_{\bf k_{1}}(t_{*})\hat \curv_{\bf k_{2}}(t_{*})\hat \curv_{\bf k_{3}}(t_{*})$] to linear order
\begin{align}\nonumber
& \langle\hat{\mathcal{R}}_{\bf k_{1}}(t_{*})\hat{\mathcal{R}}_{\bf k_{2}}(t_{*})\hat{\mathcal{R}}_{\bf k_{3}}(t_{*})\rangle = \\& 2\Re\left[ -i\int^{t_{*}}_{-\infty} dt \langle \hat{\mathcal{R}}_{\bf k_{1}}(t_{*})\hat{\mathcal{R}}_{\bf k_{2}}(t_{*})\hat{\mathcal{R}}_{\bf k_{3}}(t_{*})H_{I}(t)\rangle\right],
\label{eqn:bispec}
\end{align}
which can be evaluated in terms of the 2-point correlation function or classical 
modefunctions via Wick's theorem by making use of the unequal time correlator
\begin{align}
\langle \hat{\curv}_{\bf k}(t_1)\hat{\curv}_{\bf k'}(t_2)\rangle = (2\pi)^3\delta^{3}({\bf k}+{\bf k}')\curv_k (t_1)\curv^*_k(t_2).
\end{align}
Note that we normalize the modefunctions to have dimensions of the square root of the power spectrum
\begin{align}
\langle \hat{\curv}_{\bf k}(t_*)\hat{\curv}_{\bf k'}(t_*)\rangle = (2\pi)^3\delta^{3}({\bf k}+{\bf k}') P_\curv(k).
\end{align}
We also use the dimensionless power spectrum 
\begin{equation}
\Delta_\curv^2(k) \equiv \frac{k^3}{2\pi^2} P_\curv(k),
\end{equation}
which is more convenient for expressing dimensionless quantities.

It therefore suffices to calculate the interaction Hamiltonian to cubic order in the curvature
perturbation \cite{Weinberg:2005vy, Adshead:2008gk} $H_I \approx - \int d^3 x{\cal L}_3$ where the Lagrangian density for the curvature field ${\cal L} = {\cal L}_2 + 
{\cal L}_3 + \ldots$.   
The task of computing the bispectrum therefore begins with examining the relevant terms in the third order Lagrangian or action.

\subsection{Cubic Action}\label{sec:cubicaction}

In order to compute the bispectrum, we require the cubic action. 
We also seek to express its form in a way that the relationship between squeezed bispectrum triangles and the power
spectrum is manifest.

For a general single field model of inflation, in comoving gauge with gravitational-wave fluctuations set to zero, the cubic action is given by \cite{Chen:2006nt}
\begin{align} \label{3action}\nonumber
\action_3  = & \int dt\, d^3x  \Bigg[ \frac{a^3 \epsilon_H}{c_s^2 } \Xi
 \frac{\dot\curv^3}{H}
%
 - 2 \frac{a \epsilon_H}{c_s^2} \dot \curv  \partial_i \curv   \partial_i \chi 
\nonumber\\ & \nn
 +\frac{a^3 \epsilon_H}{c_s^4} \(\epsilon_H - 3 + 3 c_s^2 
\) \curv \dot\curv^2 \\ \nn
 & + \frac{a \epsilon_H}{c_s^2} \( \epsilon_H - 2 \sigma_1 + 1 -c_s^2  \) \curv  \( \partial \curv \)^2\\ &+ 
\frac
{a^3 \epsilon_H} {c_s^2}\frac{d}{dt}\( \frac{\epsilon_H - \eta_H}{c_s^2} \) \curv^2\dot\curv\nonumber\\&
- \frac{d}{dt} \( \frac{a^3 \epsilon_H}{c_s^2}\frac{ \epsilon_H - \eta_H}{c_s^2} \curv^2 \dot\curv \) \Bigg],
\end{align}
where \begin{align}
	\partial^2 \chi =&  \frac{a^2 \epsilon_H}{c_s^2}\dot \curv.
\end{align}
Here we have written the term usually associated with a field redefinition as a boundary term \cite{Adshead:2011bw, Arroja:2011yj}, and dropped terms that are suppressed by additional powers of $\epsilon_H$. Note that we have also dropped  boundary terms that decay at late times in standard inflation, and thus do not contribute to the bispectrum. However, we shall see that the boundary terms we have retained are in fact the most important contributions to the bispectrum in the squeezed limit.

The 
$\Xi$ term is associated with an operator $\dot \curv^3$ that
does not contribute to squeezed bispectrum configurations.   Its value depends 
on the specific model of inflation, and we have maintained generality here in the
spirit of effective field theory.\footnote{We do not consider here effective field theory terms 
associated with the extrinsic curvature which appear mainly in ghost inflation \cite{Cheung:2007st}, Galileon interactions \cite{Burrage:2010cu} or G-Inflation \cite{Kobayashi:2010cm, Kobayashi:2011nu}.}
For example in DBI inflation $\Xi=0$. In what follows in \S \ref{sec:DBI}, we shall use DBI to illustrate our technique and so we defer consideration of this term to Appendix \ref{sec:Rdot3term}.
Furthermore the $\chi$ term only contributes at ${\cal O}(\epsilon_H)$ to the reduced bispectrum and vanishes for squeezed configurations.  For completeness, we consider its effect 
in Appendix \ref{sec:chiterm}. In the remainder of this section, we therefore drop these two terms as neither play a role in  establishing the consistency of the squeezed bispectrum and power spectrum spectral index.

While the remaining four terms in the action of Eq.\ (\ref{3action}) are otherwise complete, 
each contains $c_s^{-2}$ enhanced terms.  In particular the $\curv^2 \dot \curv$ terms do, and they
contribute to squeezed bispectrum configurations.    On the other hand, we know that the theory
satisfies the consistency relation 
\begin{align}\label{eqn:consrel}
 \frac{12}{5}f_{\rm NL}   & \equiv
\lim_{k_{S}\to 0}\frac{B_{\curv}(k_S, k_{L}, k_{L})}{P_\curv(k_S)P_\curv(k_L)} \nonumber\\
& = -\frac{d\ln k_L^3 P_\curv(k_L)}{d\ln k_L}  \equiv 1-n_s
\end{align}
and so cannot have $c_s^{-2}$ enhanced terms that contribute to squeezed triangles.   While it is well known
that the consistency relation is satisfied in slow roll by the cancellation of terms, beyond slow-roll it
is difficult to establish in this form.

Our strategy is to combine the $c_s^{-2}$ terms that form total derivatives of quantities which vanish outside of
the horizon and hence provide no contribution to the bispectrum.    
We begin by  using the equation of motion (\ref{eqn:EOM}) and dropping a total space
derivative to show
\begin{equation}
 2 F \curv {\cal L}_2 = \frac{d}{dt} \left( F \frac{a^3 \epsilon_H}{c_s^2} \curv^2 \dot\curv \right)
 - \dot F\frac{a^3 \epsilon_H}{c_s^2} \curv^2 \dot\curv,
 \label{eqn:identity1}
 \end{equation}
 for an arbitrary function of time $F(t)$. Hence with $F = (\epsilon_H-\eta_H)/c_s^2$
\begin{align}\label{eqn:S3v2}\nn
\action_{3}  =&   \int dt d^3x  \curv \Bigg[\frac{\epsilon_H-\sigma_1}{c_s^2} \mathcal{H}_2 - \frac{1-c_{s}^2}{c_{s}^2}\(\mathcal{H}_2+2\mathcal{L}_2\)\\ & -2 \frac{\epsilon_H - \eta_H -\sigma_1/2}{c_s^2}\mathcal{L}_2 \Bigg].
\end{align}
Here $\mathcal{H}_2$ is the quadratic Hamiltonian density \begin{align}\label{eqn:H2}
\mathcal{H}_2 = & \frac{a^3 \epsilon_H}{c_{s}^2}\left[ \dot\curv^2 + \frac{c_s^2}{a^2}\(\partial \curv\)^2\right],
\end{align}
and the quadratic Lagrangian density was given in Eq.~(\ref{eqn:L2}).

Next we note that several terms in $\action_3$ can be grouped into a total derivative
generalizing Ref.~\cite{Creminelli:2011rh}
\begin{eqnarray}
\frac{1}{F}\frac{d}{ dt} \left(  \frac{F\curv {\mathcal{H}_2} }{H} \right)&=& \frac{\dot \curv}{H}\mathcal{L}_2 -\curv (\mathcal{H}_2+2\mathcal{L}_2) 
\nonumber\\
&&
 - 2\(\epsilon_H-\eta_H -\frac{\sigma_1}{2}\) \curv \mathcal{L}_2 
\nonumber\\
&&
+ \Big(\epsilon_H+\sigma_1+ \frac{\dot F}{F H} \Big) \curv \mathcal{H}_2  ,
\label{eqn:identity2}
\end{eqnarray}
where again $F$ is an arbitrary function of time.   With $F=1/c_s^2$, 
  \begin{align}\label{eqn:S3v5}
\action_{3} = & \int dt d^3x \Bigg[\frac{d}{dt}\(\curv\frac{\mathcal{H}_2 }{Hc_{s}^2}\)+\curv\(\mathcal{H}_2+2\mathcal{L}_2\) -\frac{\dot{\curv}}{Hc_{s}^2}\mathcal{L}_{2}\Bigg].
\end{align}
The total derivative is irrelevant here, and we can drop it.\footnote{Recall that $\mathcal{H}_2$ generates time translations for $\curv$, and so this operator results in terms which involve $\dot\curv(t_*)$ which is exponentially decaying at late times when all modes of interest are outside the horizon (at least in standard inflationary scenarios). Since the resulting correlation functions vanish exponentially fast at late times, they are irrelevant as long as we only interested in correlations late in inflation.}
Although this form for $\action_3$ efficiently groups the terms into $c_s^{-2}$ enhanced an non-enhanced
terms, it still obscures the consistency relation from the latter by not manifestly scaling with slow-roll
parameters.   
To expose this relation, we can reverse the above operations on the $\mathcal{H}_2+2\mathcal{L}_2$ term
by using the identities Eq.~(\ref{eqn:identity2}) with $F=1$ and (\ref{eqn:identity1}) with $F= 2\epsilon_H-\eta_H +\sigma_1/2$ to obtain
\begin{align}\label{eqn:actionmanip}
\action_{3} = & \int dt\, d^3x \Bigg\{
  \frac{a^{3}\epsilon_H}{c_{s}^2}\frac{d}{dt}\(2\epsilon_H - \eta_H + \frac{\sigma_{1}}{2}\) \curv^{2}\dot\curv \\  \nn& -\frac{d}{dt}\left[ \frac{a^{3}\epsilon_H}{c_{s}^2}\(2\epsilon_H - \eta_H + \frac{\sigma_{1}}{2}\) \curv^{2} \dot \curv \right] \\ 
&+(\epsilon_H +\sigma_1)\curv( \mathcal{H}_2+2\mathcal{L}_2)  +\(1-\frac{1}{c_{s}^2}\)\frac{\dot{\curv}}{H}\mathcal{L}_{2}
\Bigg\}, \nonumber
\end{align}
where we have again dropped the total derivative term for the same reasons as above. 
Note that for canonical fields $c_s=1$ and this form reproduces the $\eta_H$ dependence of the
cubic action used in  Ref.~\cite{Adshead:2011bw}.  It furthermore restores $\epsilon_H$ terms which typically do not contribute significantly to the bispectrum but complete the consistency relation in the slow-roll limit. 

In fact the slow-roll consistency relation can be trivially demonstrated given this form for $\action_3$ by making use of the technique of  Ref.~\cite{Creminelli:2011rh}.
  Notice that, to leading order in slow roll, the only term that contributes to the bispectrum in the squeezed limit is the boundary term, the second line of in Eq.\ (\ref{eqn:actionmanip}). The first term on the last line can be seen to be higher order after making use of the identity at Eq.\ (\ref{eqn:identity2}). Since the boundary term is a total derivative, we can easily evaluate its contribution to the bispectrum. Making use of the Hermiticity of the fields,  we can rewrite Eq.\ (\ref{eqn:bispec}) as the commutator
\begin{align}\nn\label{eqn:sqcommutator}
 \langle \hat{\curv}_{{\bf k}_1}\hat{\curv}_{{\bf k}_2}\hat{\curv}_{{\bf k}_3} \rangle   =&  -i\prod_{i = 1}^3 \int \frac{d^3 q_i}{(2\pi)^3} \frac{a^{3}\epsilon_H}{c_{s}^2}\(2\epsilon_H - \eta_H + \frac{\sigma_{1}}{2}\) \\ \nn& \times \left\langle \[\hat\curv_{{\bf k}_1}\hat\curv_{{\bf k}_2}\hat\curv_{{\bf k}_3} ,\hat \curv_{{\bf q}_1}\hat\curv_{{\bf q}_2}\dot{\hat\curv}_{{\bf q}_3} \]\right\rangle \\ & \times (2\pi)^3\delta^3({\bf q}_1+{\bf q}_2+{\bf q}_3) + \ldots,
\end{align}
where `$\ldots$' refers to terms that vanish in the squeezed limit. 
 We can then make use of the fact that the canonical momenta of this theory is given by,
\begin{align}
\pi_{\bf k} =  \frac{\partial \mathcal{H}_2}{\partial\dot\curv _{\bf k}} & =  2 \frac{a^3 \epsilon_H}{c_s^2} \dot\curv_{\bf k}, 
\end{align}
and satisfies the canonical commutation relation
\begin{align}
 \[\curv_{\bf k}, \pi_{{\bf k}'}\] & =  i (2\pi)^{3}\delta^{3}({\bf k} + {\bf k}'),
\end{align}
to evaluate the commutator in Eq.\ (\ref{eqn:sqcommutator}).  The leading order squeezed limit is therefore given by
\begin{align}
B_\curv(k_1,k_2,k_3) \approx & \(2\epsilon_{H} - \eta_{H} + \frac{\sigma_1}{2}\) P_\curv(k_{1})P_\curv(k_{2}) \nonumber\\
& +{\rm perm.} 
\end{align}
where ``perm." refers to the two cyclic permutations of the $k_i$ indices.  Since
in slow roll
\begin{eqnarray}
n_s - 1 &=& \frac{d \ln \Delta_\curv^2}{d\ln k} \nonumber\\
	&=& -(4\epsilon_H-2\eta_H +\sigma_1),
\end{eqnarray}
and $P_\curv(k_2) P_\curv(k_3) \ll P_\curv(k_1)P_\curv(k_2)$ for $k_1 \ll k_2 \approx k_3$, this establishes the slow-roll
consistency relation, Eq.~(\ref{eqn:consrel}).

\begin{table}[tb]
\begin{center}
\begin{tabular}{ c   c c  c c }
$i$  &  Operator &  Source & Squeezed & $T_{ij}$ Eq. 
 \\
 \hline
 &&&\\[-10pt]
0 & $\curv^2 \dot \curv$ & $2\epsilon_H -\eta_H + \dfrac{\sigma_1}{2}$ & yes & (\ref{eqn:T0}) \\[4pt]
1 & $\curv({\cal H}_2+2 {\cal L}_2)$  & $\epsilon_H + \sigma_1$ & yes & (\ref{eqn:T1})  \\[4pt]
2 & $\dot \curv {\cal L}_2$  & $\left( \dfrac{1}{c_s^2}-1 \right) \dfrac{c_s}{a H s}$ & no & (\ref{eqn:T2nm})
(\ref{eqn:T2m}) \\[10pt]
3 & $\dot \curv^3$ & $-\Xi   \dfrac{c_s}{a H s}$ & no &   (\ref{eqn:T3nm})
(\ref{eqn:T3m})\\[8pt]
4 &  $ \dot\R \( \partial\R \)  \partial \chi  $ & $\dfrac{\epsilon_H}{c_s^2}$ & no & (\ref{eqn:T4}) \\[4pt]
\end{tabular}
\end{center}
\caption{GSR bispectrum operators, sources, and triangle weights.}
\label{tab:iops}
\end{table}%

\begin{table}[tb]
\begin{center}
\begin{tabular}{    c c}
$ij$  &  $W_{ij}(x)$
 \\
 \hline
 &\\[-10pt]
00, 10, 20, 30, 40&  $x\sin x$ \\
23, 33 & $x \sin x + \cos x$ \\
01, 11, 21, 22, 24, 31, 32, 41 & $\cos x$ \\
02, 12 & $\frac{1}{x} \sin x$ \\
25, 34 & $\frac{2}{x} \sin x - \cos x$ \\
26, 35 & $W(x/2)$
\end{tabular}
\end{center}
\caption{GSR window functions for various operators $i$ and $x$-weight factors with $W$ as the power spectrum window of Eq.~(\ref{eqn:powerwindow}).}
\label{tab:windows}
\end{table}%

\begin{table}[tb]
\begin{center}
\begin{tabular}{    c c c}
$T_{ij}$ & Equilateral & Squeezed \\
\hline
 &\\[-10pt]
 $T_{00}$ & $-1$ & $-1$ \\
 $T_{01}$ & $-6$ & $-2\frac{k_L}{k_S}$ \\
  $T_{02}$ & $9$ & $4\frac{k_L}{k_S}$ \\
  \hline
  $T_{10}$ & $\frac{7}{6}$ & $1$ \\
  $T_{11}$ & $\frac{17}{3}$ & $2\frac{k_L}{k_S}$ \\
  $T_{12}$ & $-\frac{9}{2}$ & $-2 \frac{k_L}{k_S}$ \\
  \hline
  $T_{20}$ & $-\frac{1}{6}$ & $0$ \\
  $T_{21}$ & $-\frac{5}{6}$ & $0$ \\
    $T_{22}$ & $-\frac{7}{6}$ & $0$ \\
  $T_{23}$ & $\frac{5}{6}$ & 1 \\
   $T_{24}$ & $\frac{1}{4}$ & $0$ \\
    $T_{25}$ & $-\frac{37}{4}$ & $-4 \frac{k_L}{k_S}-3$ \\
    $T_{26}$ & $\frac{47}{3}$ & $\frac{8}{3} \left( \frac{k_L}{k_S} \right)^2 + 4 \frac{k_L}{k_S} + \frac{14}{3}$\\
    $T_{2B}+{\rm perm.}$ & $-\frac{15}{2}$ & $-\frac{8}{3}  \left( \frac{k_L}{k_S} \right)^2 -\frac{8}{3}$ \\
    \hline
    $T_{30}$ & $-\frac{1}{9}$ & $0$ \\
    $T_{31}$ & $-\frac{1}{3}$ & $0$ \\
    $T_{32}$ & $-\frac{2}{9}$ & $0$ \\
    $T_{33}$ & $1$ & $\frac{3}{2}$ \\
    $T_{34}$ & $-11$ & $-6\frac{k_L}{k_S}-\frac{9}{2}$\\
    $T_{35}$ & $16$ & $4 \left( \frac{k_L}{k_S} \right)^2  + 6 \frac{k_L}{k_S} + 3$ \\
    $T_{3B}+{\rm perm}$ & $-6$ & $-4 \left( \frac{k_L}{k_S} \right)^2 $  \\
    \hline
        $T_{40}$ & $\wh{\frac{1}{3}}$ & $0$\\
    $T_{41}$ &$\frac{4}{3}$  & $0$\\
      \end{tabular}
    \end{center}
    \caption{GSR triangle weights for equilateral and squeezed configurations 
    $+{\cal O}(k_S/k_L)$.  Squeezed contributions from $T_{2m}$ and $T_{3m}$, $m\ge 3$
    cancel due to Eq.~(\ref{eqn:Tcancel2}) and (\ref{eqn:Tcancel3}). }
    \label{tab:Tweights}
\end{table}%

\section{Generalized Slow Roll}
\label{sec:GSR}

In this section we construct an efficient integral formulation of  the bispectrum contributions from Eq.~(\ref{eqn:bispec}) and the cubic action of Eq.~(\ref{eqn:actionmanip}) that allows arbitrary time
variation in the slow-roll parameters $\epsilon_H, \eta_H, \sigma_1$.
The remaining model-dependent $\dot \curv^3$ term and $\epsilon_H$ suppressed terms are  considered in Appendix \ref{sec:Rdot3term} and \ref{sec:chiterm} respectively.   

\subsection{Formalism}

To evaluate the bispectrum exactly, we need to solve for the background evolution for the slow-roll parameters
$\epsilon_H$, $\eta_H$, $\sigma_1$ and sound speed $c_s$ as well as the modefunctions $\curv_k$
for each mode in the triangle configuration.   The modefunctions themselves are dependent on
the slow-roll parameters and beyond the slow-roll approximation, where the slow-roll parameters
are taken to be constant, there is no general analytic
solution for their behavior.    

The generalized slow-roll approach (GSR) \cite{Stewart:2001cd,Choe:2004zg,Dvorkin:2009ne}
provides an iterative approximation to the modefunctions.  Their equation of motion Eq.~(\ref{eqn:EOM})  can be recast as \cite{Hu:2011vr}
\begin{align}\label{eqn:yeqn}
        \frac{d^2y}{dx^2} + \left(1 - \frac{2}{x^2} \right) y = \frac{g(\ln s)}{x^2}y,
\end{align}
where 
\begin{equation}
y \equiv \sqrt{\frac{k^3}{2\pi^2} } \frac{f}{x} \curv_k,
\end{equation} 
$x= k s$, and
the sound horizon
\begin{equation}
s(t) = \int_t^{t_{\rm end}} \frac{c_s dt}{a}  ,
\end{equation}
with $t_{\rm end}$ defining the end of inflation.
Here
\begin{align}
        g \equiv \frac{f'' - 3 f'}{f},
\end{align}
with  $' \equiv d/d\ln s$ throughout and
\begin{align}\label{eqn:fdef}
	f^2 & = 8 \pi^2 \frac{\ep \cs}{H^2} \esq \frac{a H s}{\cs} \dir^2.
\end{align}
Note that in the slow-roll limit
$
\Delta^2_{\curv} \approx f^{-2}
$
(see Eq.~\ref{eqn:power1}).
 In the GSR approximation, one first defines the solution to Eq.~(\ref{eqn:yeqn}) with
$g=0$ and Bunch-Davies initial conditions
\begin{equation}
y_0(x) = \left( 1 + {i \over x} \right) e^{i x} ,
\end{equation}
and then  replaces the RHS of Eq.~(\ref{eqn:yeqn}) with $y \rightarrow y_0$.     The solution 
to first order in $g$ is
\begin{align}\label{eqn:firstordermode}
y(x) = y_{0}(x) -\int_{x}^{\infty}\frac{d u }{u^2}
g(\ln  s)y_0(u)\Im[y^*_{0}(u)y_{0}(x)],
\end{align}
where $u= k s$.  With these relations we can define an integral approximation to the bispectrum
to leading order in the slow-roll deviations $g$, $\epsilon_H$, $\eta_H$ and $\sigma_1$.
For $\action_3$ operators that already include slow-roll parameters only the zeroth order
$y_0$ modefunctions are required whereas those
that involve none, i.e.~$\dot \curv {\cal L}_2$ for DBI inflation, the first-order modefunction 
correction contribute to first order in the GSR approximation.

\subsection{Integral Form}

All contributions from the $\action_3$ operators can be cast into integral form for the
dimensionless bispectrum
\begin{equation}
\frac{\cal G}{k_1 k_2 k_3} = \frac{k_1^2 k_2^2 k_3^2}{(2\pi)^4  A_s^2} B_\curv(k_1,k_2,k_3),
\end{equation}
where $ A_s$ is a constant of order the dimensionless power spectrum $\Delta_\curv^2 = k^3 P_\curv/2\pi^2$.  In the leading order GSR approximation
these integrals depend only on the perimeter of the triangle $K=k_1+k_2+k_3$ rather
than its shape.   Thus this description enables a highly efficient computation of all bispectrum
triangles from a handful of one dimensional integrals.

We will group our integral results according to 
sources  $S_{ij} (\ln s)$ indexed with: $(i)$ the operators they correspond to;  $(j)$ the scale $x= K s$ at which
the operator sources contribute.  The integrals are given by
 \begin{equation}
 I_{ij}(K) = S_{ij}(\ln s_*) W_{ij}(K s_*) + \int_{s_*}^\infty \frac{ds}{s}  S_{ij}'(\ln s) W_{ij}(K s)
 \label{eqn:Iform}
 \end{equation}
 where $W_{ij}$ are fixed window functions that are independent of the source.
The triangle shape dependence is carried by $T_{ij}$ which are universal functions
of $(k_1,k_2,k_3)$ such that any bispectrum triangle can be computed as
\begin{eqnarray}
\frac{\cal G}{k_1 k_2 k_3} &=& \frac{\Delta_\curv(k_1) \Delta_\curv(k_2) \Delta_\curv(k_3)}{4 A_s^2} \Big\{  \sum_{ij} T_{ij} I_{ij}(K)  \nonumber\\ && + [
T_{2B} I_{26}( 2 k_3) +{\rm perm.}] \Big\}.
\label{eqn:GSRBi}
\end{eqnarray}
In the following sections, we consider the contribution of each operator composing $\action_3$ in turn.  We
summarize these terms in Tab.~\ref{tab:iops}, their associated windows in Tab.~\ref{tab:windows}, and limiting cases of the triangle weights in Tab.~\ref{tab:Tweights}.

In the leading order GSR approximation, the power spectrum itself is given by 
the integral \cite{Hu:2011vr}
\begin{align}
\label{eqn:GSRpower}
\ln \Delta_\curv^{2} &= G(\ln s_{*})W(k s_*) + \int_{s_{*}}^\infty {d s\over s} W(ks) G'(\ln s),
\end{align}
where the power spectrum window function is
\begin{equation}
W(u) = {3 \sin(2 u) \over 2 u^3} - {3 \cos (2 u) \over u^2} - {3 \sin(2 u)\over 2 u} 
\label{eqn:powerwindow}
\end{equation}
and the source function
\begin{equation}
G = - 2 \ln f    + {2 \over 3} (\ln f )'   .
\end{equation}
This power spectrum expression then completes the form of the bispectrum in Eq.~(\ref{eqn:GSRBi}).
 
\subsubsection{$i=0$: $\curv^2 \dot \curv$}
\label{sec:R2Rdot}

We start by considering the first two terms in Eq.~(\ref{eqn:actionmanip}) which involve the
 $\curv^2 \dot \curv$ operator.   This operator was considered in detail in 
Ref.~\cite{Adshead:2011bw} for canonical $(c_s=1)$ scalar fields and the results carry directly over
to the general case with the source replacement
\begin{equation}
S_{00} = S_{01} = S_{02} =  \frac{1}{f}\left( 2\epsilon_H - \eta_H + \frac{\sigma_{1}}{2} \right).
\label{eqn:S0source}
\end{equation}	
Namely the windows
\begin{eqnarray}
W_{00} (x) &=&   x \sin x,  \nonumber\\
W_{01} (x) &=& \cos x,  \nonumber \\
W_{02}(x) &=&   \frac{\sin x}{x},
\end{eqnarray}
and the triangle weights
\begin{eqnarray}
T_{00} &=& -1  , \nonumber\\
T_{01} &=& - \frac{ \sum_{i\ne j} k_i k_j^2}{ k_1 k_2 k_3} ,\nonumber\\
T_{02} &=& \frac{K\sum_i k_i^2}{k_1k_2k_3} ,
\label{eqn:T0}
\end{eqnarray}
are identical to the canonical case.   Compared with the treatment in Ref.~\cite{Adshead:2011bw},  here we include the $\epsilon_H$ correction associated with
the full $\curv^2 \dot \curv$ operator in Eq.~(\ref{eqn:actionmanip}).

As in the canonical case, the appearance of $1/f$ in the
source makes the integrals in Eq.~(\ref{eqn:Iform}) involve total derivatives $S_{ij}'$ and hence guarantees
that the bispectrum remains constant once all 3 $k$-modes have exited the horizon.   It was shown 
in Ref.~\cite{Adshead:2011bw} that this is the consequence of first-order modefunction corrections using
Eq.~(\ref{eqn:firstordermode}).

Note that both the $T_{01}$ and $T_{02}$ terms contribute to squeezed triangles $k_1=k_S \ll k_L = k_2 \approx k_3$
\begin{eqnarray}
\lim_{k_S\ll k_L}  T_{02} = -2 T_{01}=  4\frac{k_L}{k_S}
\end{eqnarray}
and hence are involved in establishing the consistency
relation. 

\subsubsection{$i=1$: $ \curv({\cal H}_2+2 {\cal L}_2$)}
\label{sec:H2L}

The $\curv({\cal H}_2+2 {\cal L}_2)$ can likewise be calculated with the GSR expansion.   Again only the
zeroth order modefunctions are required throughout and constancy of the bispectrum on superhorizon
scales is automatic due to an integration by parts which brings the source into the form
\begin{equation}
S_{10} = S_{11}= S_{12}  =\frac{\epsilon_H + \sigma_1}{f},
\end{equation}	
with windows
\begin{eqnarray}
W_{10}(x) &=&    x \sin x , \nonumber\\
W_{11}(x) &=&   \cos x ,\nonumber \\
W_{12}(x) &=&   \frac{\sin x }{x},
\end{eqnarray}
and triangle weights
\begin{eqnarray}
T_{10} &=&  \frac{3}{2} - \frac{\sum_i k_i^2}{K^2}, \nonumber\\
T_{11}
&=& \frac{1}{k_1 k_2 k_3} \Big[  \frac{1}{2} \sum_{i\ne j} k_i k_j^2  + \frac{4}{ K} \sum_{i>j} k_i^2 k_j^2 - \frac{2}{K^2} \sum_{i\ne j} k_i^2 k_j^3  
\Big] ,
\nonumber\\
T_{12} &=&  -\frac{K \sum_i k_i^2}{2 k_1 k_2 k_3}. 
\label{eqn:T1}
\end{eqnarray}
Both $T_{11}$ and $T_{22}$ contribute in the squeezed limit and in fact
\begin{eqnarray}
\lim_{k_S\ll k_L} T_{11} = -T_{12} = 2\frac{k_L}{k_S}.
\end{eqnarray}
In the slow-roll limit only the boundary terms in $I_{11}$ and $I_{12}$ contribute since source
derivatives involve evolution in the slow-roll parameters (see Appendix \ref{app:sr} for further discussion).   Given that $W_{11}(0)=W_{12}(0)=1$, the
sum of the two terms vanishes and the
operator does not contribute to squeezed triangles in the slow-roll limit.    They do contribute beyond the
slow-roll limit where the evolution of the sources enters.
 
\subsubsection{$i=2$: $\dot \curv {\cal L}_2$}
\label{sec:dotRL}

The only operator that is enhanced by $c_s^{-2}$ is  $\dot \curv {\cal L}_2$.  This one
is also special in that it is not suppressed by any slow-roll parameters.     To calculate its
bispectrum consistently to leading order in the GSR approximation, we must use the first
order modefunction expansion from Eq.~(\ref{eqn:firstordermode}) [see \S \ref{sec:Rdot3term} for more details on an analogous calculation].  These terms contribute comparably to those that describe the evolution of the $c_s^{-2}$ enhanced term itself
\begin{equation}
S_{2}(\ln s) =
\left( \frac{1}{c_s^2}-1 \right)\left( \frac{c_s}{a H s} \right) \frac{1}{f},
\label{eqn:S2}
\end{equation}
using zeroth order modefunctions $y_0$.  

For the latter effect, it is computationally advantageous to isolate the $S_2$ evolution terms by integrating
by parts so that the new sources are
\begin{eqnarray}
S_{20} &=& S_{21} = S_2' ,\nonumber\\
S_{22} &=& S_2,
\end{eqnarray}
with
\begin{eqnarray}
T_{20} 
&=& \frac{\sum_i k_i^2 - 2 \sum_{i>j} k_i k_j}{2 K^2}, 
\label{eqn:T2nm} \\
T_{21}  &=& 
 \frac{1}{k_1 k_2 k_3} \Big[  \frac{1}{2} \sum_{i\ne j} k_i k_j^2  - \frac{6}{ K} \sum_{i>j} k_i^2 k_j^2 + \frac{4}{K^2} \sum_{i\ne j} k_i^2 k_j^3  
\Big] 
\nonumber\\
&& - \frac{1}{2} ,\nonumber\\
 T_{22} &=& 
\frac{1}{k_1 k_2 k_3} \Big[ \frac{1}{2}\sum_i k_i^3 - \frac{4}{K} \sum_{i>j} k_i^2 k_j^2 + \frac{2}{ K^2} \sum_{i\ne j} k_i^2 k_j^3  \Big], \nonumber
\end{eqnarray}
and
\begin{eqnarray}
W_{20}(x) &=& x \sin x, \nonumber\\
W_{21}(x)   &=& W_{22}(x)= \cos x .
\end{eqnarray}
By constructing the integrals in this manner, we guarantee that evolution in $S_2$ that has compact
support in $S_2''$ results in rapidly convergent integrals as $x \rightarrow \infty$ which do not
require regulation.  This should be compared with the unmanipulated $S_2$ integrals or the
exact integration over modefunctions, both of which have window or modefunction weights
that diverge as $x^3$ (see \S \ref{sec:Rdot3term}).

Note that the boundary term of the $I_{22}$  integral gives the well known result that 
\begin{equation}
{\cal G} \approx \left( \frac{1}{c_s^2}-1\right) \Big[ \frac{1}{8}\sum_i k_i^3 - \frac{1}{K} \sum_{i>j} k_i^2 k_j^2 + \frac{1}{ 2K^2} \sum_{i\ne j} k_i^2 k_j^3  \Big]
\end{equation}
to zeroth order in slow-roll parameters.  In Appendix \ref{app:sr}, we derive an expression that
is correct to first order in the slow-roll approximation.

Unlike the other operators, there are leading order contributions associated with the deviation of the modefunctions from their $y_0$ de Sitter form.    The nested 
integration of the modefunction correction inside the bispectrum integral can be unwound by integration
by parts, using the leading order approximation of $S_2$=const., leaving a new source 
\begin{equation}
g S_2 = S_{23}'=S_{24}'=S_{25}'=S_{26}'  ,
\label{eqn:S2m}
\end{equation}
with
\begin{eqnarray}
T_{23}
 &=& \frac{3}{2} -\frac{2\sum_{i>j} k_i k_j }{K^2},\nonumber\\
T_{24} &=& \frac{(K-2k_1)(K - 2k_2)(K-2k_3)}{4 k_1 k_2 k_3},
\nonumber\\
T_{25} &=& -\frac{1}{8 k_1 k_2 k_3 (K-2k_3) K^2} \nonumber\\&& \times \Big[ 
(k_1^2-k_2^2)^2 (k_1^2 + 6 k_1 k_2 + k_2^2) \nonumber\\&&
+ 4 (k_1-k_2)^2 (k_1+k_2) (k_1^2 + 6 k_1 k_2 + k_2^2) k_3  \nonumber\\&&
+ 2 (3 k_1^4 + 23 k_1^3 k_2 + 64 k_1^2 k_2^2 + 23 k_1 k_2^3 + 3 k_2^4)k_3^2 \nonumber\\&&
+ 16 k_1 k_2 (k_1 + k_2) k_3^3 
- (7 k_1^2 + 20 k_1 k_2 + 7 k_2^2) k_3^4  \nonumber\\&&
- 4(k_1 + k_2) k_3^5 \Big] + {\rm perm.},\nonumber \\
T_{26} &=& \frac{1}{12 k_1 k_2 k_3 (K-2k_3)^2 K^2} \nonumber\\&& \times  \Big[ 
 (k_1-k_2)^2(k_1+k_2)^3 (k_1^2+ 6 k_1 k_2 + k_2^2) \nonumber\\&&
+3(k_1^2-k_2^2)^2 (k_1^2 + 6 k_1 k_2 + k_2^2) k_3 \nonumber\\&&
+ 2(k_1+k_2)(6 k_1^4 + 35 k_1^3 k_2 + 106 k_1^2 k_2^2 
\nonumber\\&&\quad
+ 35 k_1 k_2^3 + 6 k_2^4) k_3^2 \nonumber\\&&
+ 2(2 k_1^4 + 5 k_1^3 k_2 - 26 k_1^2 k_2^2 + 5 k_1 k_2^3 + 2 k_2^4) k_3^3\nonumber\\&&
-(k_1 + k_2)(19 k_1^2 + 44 k_1 k_2 + 19 k_2^2) k_3^4  \nonumber\\&&
+(-9 k_1^2 + 4 k_1 k_2 - 9 k_2^2) k_3^5 \nonumber\\&&
+ 6 (k_1 + k_2) k_3^6 + 2 k_3^7 \Big] + {\rm perm.} 
\label{eqn:T2m}
\end{eqnarray}
Here ``perm" means the 2 additional cyclic permutations.
Integration by parts on the modefunction expansion also leaves a boundary term that is described by
\begin{eqnarray}
T_{2B} &=& \frac{1}{6k_1 k_2 k_3 (K-2 k_3)^2}  \\&& \times
 \Big[
( k_1-k_2)^2 (k_1 + k_2) (k_1^2 + 3 k_1 k_2 + k_2^2) 
\nonumber\\&&
- 2 (k_1-k_2)^2 ( k_1^2 + 3 k_1 k_2 + k_2^2) k_3 \nonumber\\&&
- (k_1 + k_2) (3 k_1^2 + 5 k_1 k_2 + 3 k_2^2) k_3^2 \nonumber\\&&
+ (3 k_1^2 - 2 k_1 k_2 + 3 k_2^2) k_3^3 
+ 2 (k_1+k_2) k_3^4 - k_3^5 
\Big], \nonumber
\end{eqnarray}
and $I_{26}(2k_3)$ in Eq.~(\ref{eqn:GSRBi}).
The windows associated with these terms are
\begin{align}
W_{23}(x) =& x \sin x + \cos x, \\
W_{24}(x) =& \cos x ,\nonumber\\
W_{25}(x) =& 2\frac{\sin x}{x} - \cos x, \nonumber\\
W_{26}(x) =& 12\left(  \frac{\sin x}{x^3} - \frac{\cos x}{x^2} -\frac{\sin x}{ 4 x}\right).\nonumber
\end{align}
Note that $W_{26}(x) = W( x/2)$, the window function of the GSR power spectrum defined in 
Eq.~(\ref{eqn:powerwindow}).

While the evaluation at $s_*$  in Eq.~(\ref{eqn:Iform}) 
for these 5 modefunction terms would formally require integrating
$g S_2 $, the sum exactly vanishes for all triangles  and $K s_* \ll 1$. They may be omitted in practice so
long as the same $s_*$ is taken for each.  This cancellation is a consequence of
the triangle weights obeying
\begin{equation}
\sum_{j=3}^6 T_{2j} + (T_{2B} + {\rm perm.}) = 0.
\label{eqn:Tcancel2}
\end{equation}

 Likewise, while individual terms would
seem to contribute to squeezed triangles (see Tab.~\ref{tab:Tweights}), the sum is suppressed by $k_S/k_L$ for any source evolution
{so long as that source contributes when $k_S s \ll 1$}. Note that the consistency relation is not expected to hold if the long wavelength mode, $k_S$, is inside the horizon since it can no longer be considered as a change in the background for the evolution of the short-wavelength, $k_L$, modes.

\subsection{Consistency Relation}
\label{sec:consistency}

With these integral expressions for the bispectrum and power spectrum, we can now re-examine the
consistency relation for squeezed bispectra, Eq.~(\ref{eqn:consrel}).   Amongst the integrals 
only $I_{01}$, $I_{02}$, $I_{11}$, $I_{12}$ contribute to squeezed triangles as $k_L/k_S$.
 Note that none of these terms are 
enhanced by $c_s^{-2}$ relative to the power spectrum.  To leading order,
\begin{eqnarray}
\label{eqn:GSRconsistency}
\frac{12}{5} f_{\rm NL}
&\approx &
-2 \frac{f'}{f}\Big|_{s_*} 
+  f_* \int_{s_*}^\infty \frac{d s}{s}   \bigg[ \left( \frac{\epsilon_H}{f} \right)' W_{\epsilon}( k_L s) \\ &&  \qquad 
+
\left( \frac{\eta_H}{f} \right)' W_{\eta}( k_L s) 
+
\left( \frac{\sigma_1}{f} \right)' W_{\sigma}( k_L s) \bigg], \nonumber
\end{eqnarray}
where
\begin{eqnarray}
 W_{\epsilon}(x) &=& -2 {\cos (2 x)} + \frac{3}{x}  \sin(2 x),\nonumber\\
 W_{\eta}(x) &=& {2 \cos (2 x)} - \frac{2}{x}  \sin(2 x), \nonumber\\
 W_{\sigma}(x) &=& {\cos(2 x)}.
 \end{eqnarray}
Here we have again evaluated the boundary term by assuming $s_*$ is an epoch during slow-roll
\begin{equation}
2 \frac{f'}{f}\Big|_{s_*} \approx - (4\epsilon_H - 2\eta_H +\sigma_1)|_{s_*} .
\end{equation}

This should be compared with the local slope of the power spectrum \cite{Adshead:2012xz}
\begin{eqnarray}
\label{eqn:slope}
\frac{ d \ln \Delta_\curv^2}{d \ln k}\Big|_{k_L}  &=& \int_{s_*}^\infty \frac{ds}{s} W'(k_L s) G'(\ln s)
\\
&=&
2 \frac{f'}{f}\Big|_{s_*} + \int_{s_*}^\infty \frac{ds}{s} \left( \frac{f'}{f} \right)' W_{n}(k_L s), \nonumber
\end{eqnarray}
where
\begin{equation}
W_{n}(x) =-2 \cos(2 x) + \frac{2}{x} \sin(2x).
\label{eqn:tiltwindow}
\end{equation}

The boundary term obviously matches between Eqs.~(\ref{eqn:GSRconsistency}) and (\ref{eqn:slope})
 and establishes the
slow-roll consistency relation.  The integral piece contributes when there are features
that violate the slow-roll approximation. 
For a sharp feature at $k_L s \gg 1$, the parameters with the highest number of derivatives of $H$ and $c_s$ dominate and  we can approximate
\begin{equation}
\left( \frac{f'}{f} \right)'  \approx f_* \left( \frac{\eta_H + \sigma_1/2}{f} \right)',
\end{equation}
which matches the $\eta_H$ term in Eq.~(\ref{eqn:GSRconsistency}) and the $\sigma_1$ term in
the $k_L s \gg 1$ limit assumed.   
Note that $\sigma_1$ enters with opposite sign relative to $\eta_H$ between the slow-roll  and
sharp feature expressions.   This is a consequence of the ${\cal H}_2 + 2 {\cal L}_2$ term entering
the latter but not the former.
For $k_L s \sim 1$, the power spectrum source $G'$ no longer
appears as a sharp function compared to the windows and so other terms that impact its shape matter 
\cite{Miranda:2012rm}.
We illustrate below that the GSR approximation maintains the consistency relation even in this region.  {Finally note that once all the 
terms are considered, including window function expansions, the contribution of terms
not involved in the consistency relation
is suppressed by ${\cal O}(k_S/k_L)^2$ independently of the sources as expected \cite{Creminelli:2011rh}.}

\section{DBI Step Feature}
\label{sec:DBI}

In this section, we illustrate the GSR integral construction of the bispectrum from \S \ref{sec:GSR} in a
DBI model with sharp features in the sound speed.    We review the DBI model  in \S \ref{sec:DBImodel},
test the GSR approximation in \S \ref{sec:DBIGSR} and discuss analytic scaling results in
\S \ref{sec:DBIanalytic}.   Appendix \ref{app:parameters} gives details on how we  set DBI parameters   that are matched to the \emph{Planck} data.

\subsection{Model}
\label{sec:DBImodel}

The DBI action is a specific incarnation of the general 
$k$-inflation  
 action  \cite{ArmendarizPicon:1999rj}
\begin{equation}
\action = \int d^4 x \sqrt{-g}\, \left[ \frac{R}{2} + P(X,\phi) \right].
\end{equation}
Here $R$ is the Ricci scalar.   
The scalar field Lagrangian is taken to be a general function of the field value $\phi$ 
and its  kinetic term 
\begin{equation}
X =- {1 \over 2} \nabla^\mu \phi \nabla_\mu\phi .
\end{equation}
The scalar field behaves as a perfect fluid with pressure $P$,
\begin{equation}
\rho = 2 X P_{,X} - P , \quad
c_s^2 = {P_{,X}  / \rho_{,X}}.
\end{equation}
In these models the $\Xi$ term in the cubic action Eq.~(\ref{3action}) is given by
\begin{equation}
 \Xi =\frac{1}{c_s^2}-1 - 2 \frac{\lambda}{\Sigma} ,
 \label{eqn:XiP}
\end{equation}
where \cite{Seery:2005wm}
\begin{align}
	\Sigma    \equiv & \frac{H^2 \epsilon_H}{c_s^2},\quad 
	\lambda  \equiv  X^2 P_{,XX} + \frac{2}{3} X^3 P_{,XXX} .
\end{align}

In the DBI case, 
\begin{equation}
P(X,\phi) = \left[
1-\sqrt{1 - 2  X/T(\phi)} \right] T(\phi)- V(\phi),
\end{equation}
where $T(\phi)$ gives the warped brane tension, and $V(\phi)$ is the interaction potential.
Note that $\Xi = 0$ and so the only terms involved in the bispectrum are those given in \S 
\ref{sec:GSR}.

We illustrate our bispectrum technique on models where $T(\phi)$ has a step feature \cite{Bean:2008na}.   Details of this model including the background evolution and construction of the slow-roll parameters
can be found in Ref.~\cite{Miranda:2012rm}; we review its basic properties below.    The warp factor
\begin{align}\label{eqn:T(phi)}
	T(\phi) = \frac{\phi^4}{\lambda_B} [1 + b F(\phi) ], 
\end{align}
has a $\tanh$ step-like feature
\begin{align}
	  F(\phi) = \tanh\left(\frac{\phi - \phi_s}{d} \right)-1 ,
\end{align}
and we consider infra-red DBI inflation \cite{Chen:2004gc,Chen:2005ad} where $\phi$ inflates on the  potential 
\begin{equation}
V(\phi) =  V_0\(1 - \frac{1}{6}\beta \phi^2\),
\end{equation}
rolling from small to large values.  We have chosen a convention that after the feature, $T(\phi)$ goes back to its $b=0$ value.   We assume that inflation ends when $\phi=\phi_{\rm end}$.  Note that the step is $2 b$ in amplitude.

The parameters $\lambda_B$, $\beta$, and $\phi_{\text{end}}$ are chosen 
to fix the amplitude and tilt of the power spectrum to the maximum likelihood
of the  \emph{Planck} data \cite{Ade:2013lta}
\begin{eqnarray}
A_s &=&   2.69 \times 10^{-9} , \nonumber\\
n_s-1 &=& -0.0381,
\label{eqn:ssplanck} 
\end{eqnarray}
as well as the sound speed
at a sound horizon of
$s=3.692$ Gpc
 in the absence of the step $(b=0)$ as described
in Appendix \ref{app:parameters}. 
  The observables are weakly dependent on $V_0$ at
fixed $s$ and so we follow Ref.~\cite{Miranda:2012rm} in choosing
\begin{align}
	V_0 &=7.10 \times 10^{-26}.
\end{align}
In the presence of a step, the power spectrum and sound speed vary from these
values but we consider $A_s$ as a fixed number and label models with $c_s$ evaluated
at $b\rightarrow 0$.

The value of the sound horizon 
 in Eq.~(\ref{eqn:ssplanck}) is motivated by the recent  \emph{Planck} results \cite{Ade:2013rta} where a
step feature at this scale
\begin{equation}
s_s\equiv s(\phi_s)= 3.692\, {\rm Gpc}
\end{equation}
with power spectrum amplitude in the DBI model of \cite{Miranda:2012rm}
\begin{equation}
{\cal C} = -2 \frac{1-c_{s}}{1+c_{s}} \frac{b}{\sqrt{1-2b}}  \approx 0.1.
\label{eqn:poweramp}
\end{equation}
Unlike the {\it  \emph{WMAP}} data \cite{Adshead:2011jq},  the  \emph{Planck} data   reveals a preference
for a finite width with their increased angular resolution \cite{Ade:2013rta} 
\begin{equation}
x_d = \frac{\phi' }{\pi d} = 87.4
\label{eqn:xd}
\end{equation}
with this set of parameters favored at  $\Delta 2\ln{\cal L} \approx 12$.  
We shall see that by preferring a specific $x_d$, the  \emph{Planck} data favor a specific
and large maximum for the bispectrum.   We therefore illustrate 
 bispectrum results with these values
for $b$, $\phi_s$ and $d$ below while also testing for robustness to parameter variations.

\subsection{GSR Tests}
\label{sec:DBIGSR}

\begin{figure}[t]
\psfig{file=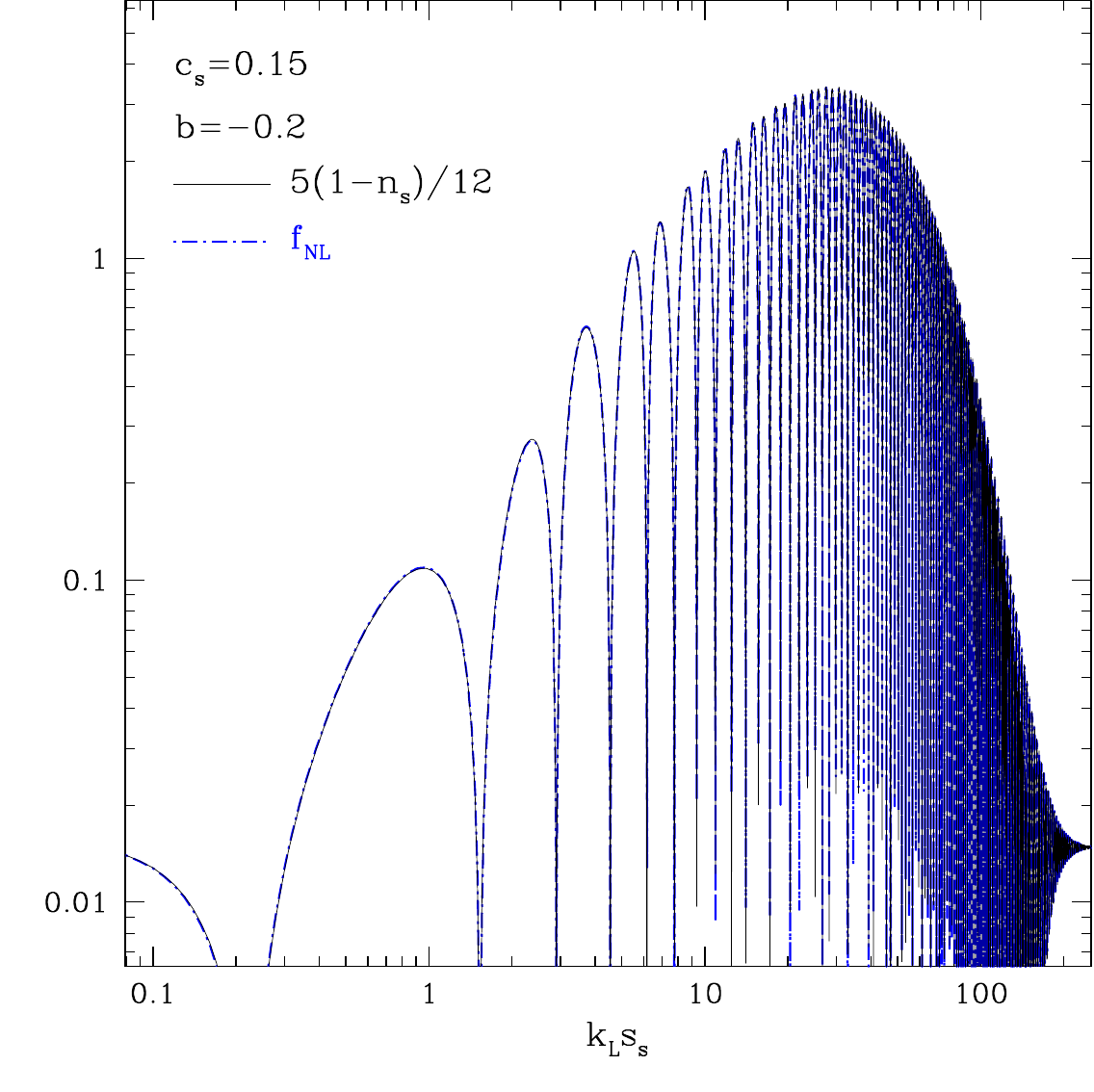, width=3.25in}
\caption{Test of the consistency relation: $n_s-1$  computed exactly from the power spectrum vs $f_{\rm NL}$ computed exactly from the bispectrum for a large amplitude step $b=-0.2$,  $x_d = 15$, $c_{s}=0.15$, $k_S/k_L = 10^{-4}$. 
}
\label{fig:Mukanov_Squeeze_big}
\end{figure}

To test the accuracy of the GSR approximation, we compare the integral approximation at  Eq.\ (\ref{eqn:GSRBi}) to a numerical computation of the full bispectrum using modefunctions obtained from numerically solving the linear equations of motion.  

We begin by testing the squeezed limit of the bispectrum which mainly checks the validity of the method.
We have constructed the GSR approximation in a manner in which the squeezed limit and its relationship to the local slope of the power spectrum is manifest in the $i=0,1$ terms
(see \S \ref{sec:consistency}).   We show in Fig.~\ref{fig:Mukanov_Squeeze_big}\
that the consistency relation itself is satisfied in the exact calculation of the
bispectrum and power spectrum.  We choose to fix the ratio $k_S/k_L =  10^{-4}$.  Since 
$k_S s_s \ll 1$ even for $k_L$ in the damping tail, agreement with the consistency
relation is expected.   We know analytically that the
$i=2$ operator contribution is suppressed by $(k_S/k_L)^2$ compared to true
squeezed contributions in this case
and so it is computationally most efficient to drop them outright.
Note that the full extent of this suppression  is achieved from cancellation of integral
terms which is difficult to reproduce numerically.   Nonetheless our numerical integration is sufficient
to make residual contributions from numerical errors negligible here compared with the true ones from
$i=0,1$.

\begin{figure}[t]
\psfig{file=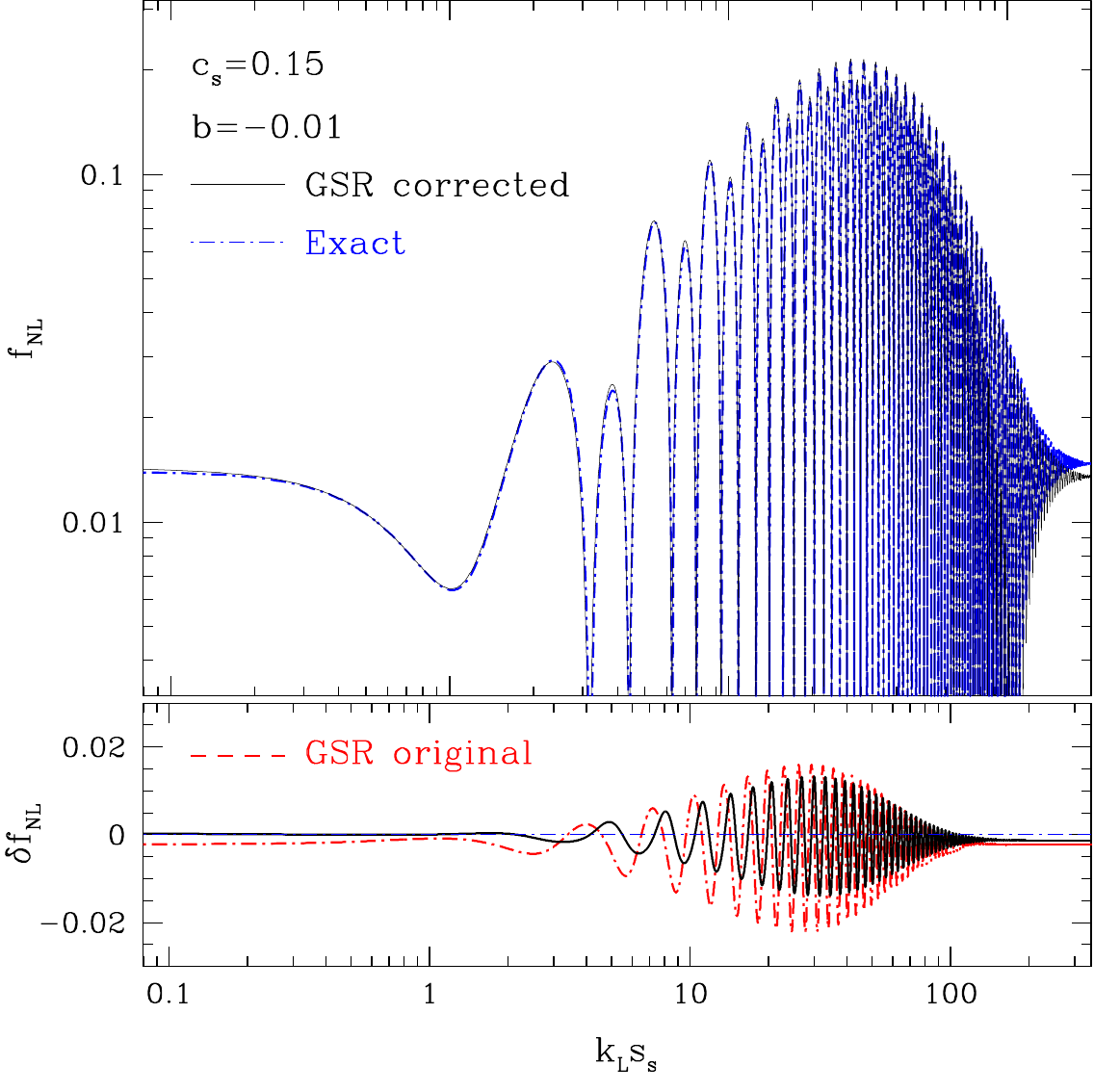, width=3.25in}
\caption{Exact vs GSR approximation for squeezed triangles for a small amplitude sharp step $b=-0.01$,  $x_d = 15$, $c_{s}=0.15$, $k_S/k_L = 10^{-4}$.
 Top panel shows the GSR result with a small correction due
to $k_S$ exiting the horizon
many efolds before the feature from Eq.~(\ref{eqn:gsr_squeeze_corr}). Bottom panel shows the error in the corrected
GSR approximation as well as that of the original GSR approximation.
After correction, the errors are mainly an out of phase component.}
\label{fig:Mukanov_GSR_small}
\end{figure}

The accuracy of the GSR approximation for these highly squeezed triangles is shown in Fig.~\ref{fig:Mukanov_GSR_small}.  In this case the approximation has small but notable amplitude errors
even in the small amplitude case.    As shown in Ref.~\cite{Adshead:2012xz}, these
errors in amplitude arise due to slow-roll corrections in the value of $f$
between when the mode $k_S$ left
the horizon and when the features at $k_L$ are imprinted.    For the step feature the
latter is fixed at $s_s$ and the correction is \cite{Adshead:2012xz}
\begin{align}\label{eqn:gsr_squeeze_corr}
	R = 1 + \frac{n_s-1}{2} \ln \(\frac{k_S s_s}{x_{f}}\), \ k_S s_s < x_{f}
\end{align}
and $R = 1$ otherwise. Here $x_{f}= e^{2-\gamma_E}/2 \approx 2.07$ is the freezeout epoch for the tilt (see Tab.~\ref{tab:freezeout}). 
In  Fig.~\ref{fig:Mukanov_GSR_small}, we also show that with this correction,  the
remaining error from the feature is in a small out-of-phase component.   These too can be corrected with the first order techniques of Ref.~\cite{Adshead:2012xz} but note that in this
example they peak at an unobservable $\delta f_{\rm NL}$ below $0.02$.
There is an additional
correction for the  slow-roll contributions on either side of the features due to
the evolution in $f$ between horizon crossing of $k_S$ and $k_L$.  Since these
are slow-roll suppressed, correcting them is never relevant.

Next we compare the equilateral bispectrum where the $c_s^{-2}$ enhanced $i=2$ term
contributes.    Here  we fix the basic parameters
as described in the previous section according the  \emph{Planck} best fit,
but allow the sound speed and amplitude $b$ of the feature to vary.

\begin{figure*}[t]
\psfig{file=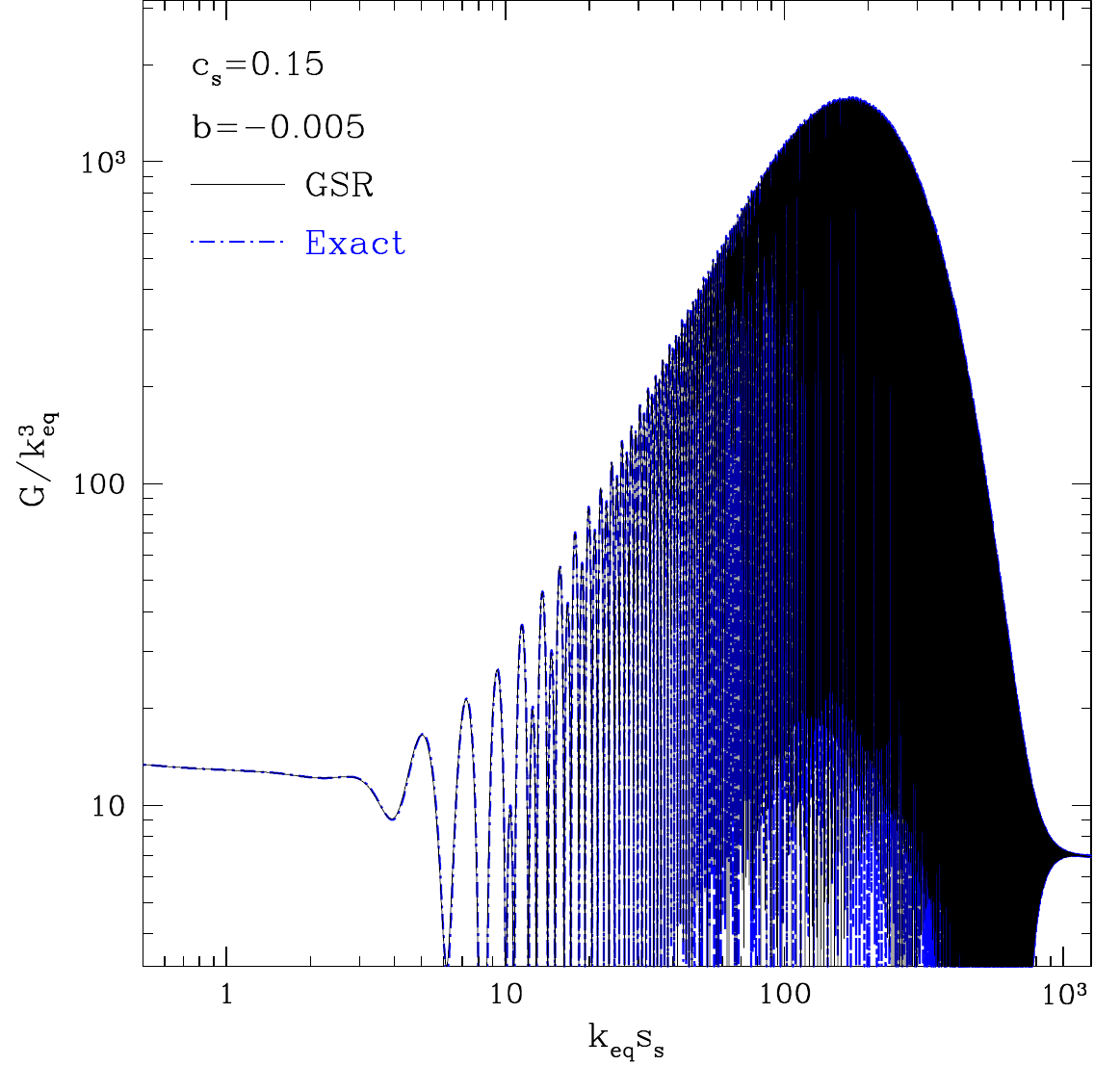, width=3in}
\psfig{file=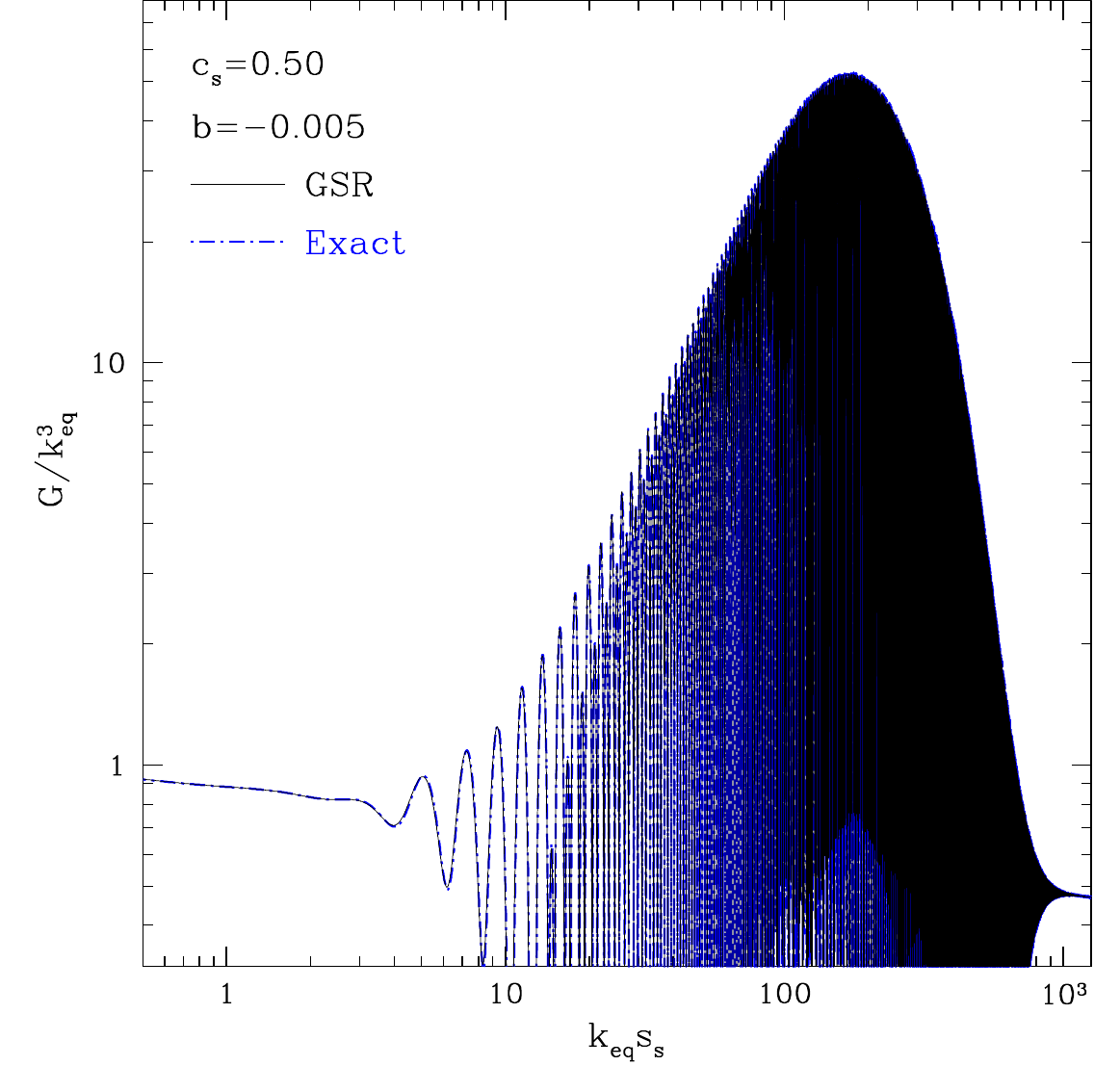, width=3in} \\
\psfig{file=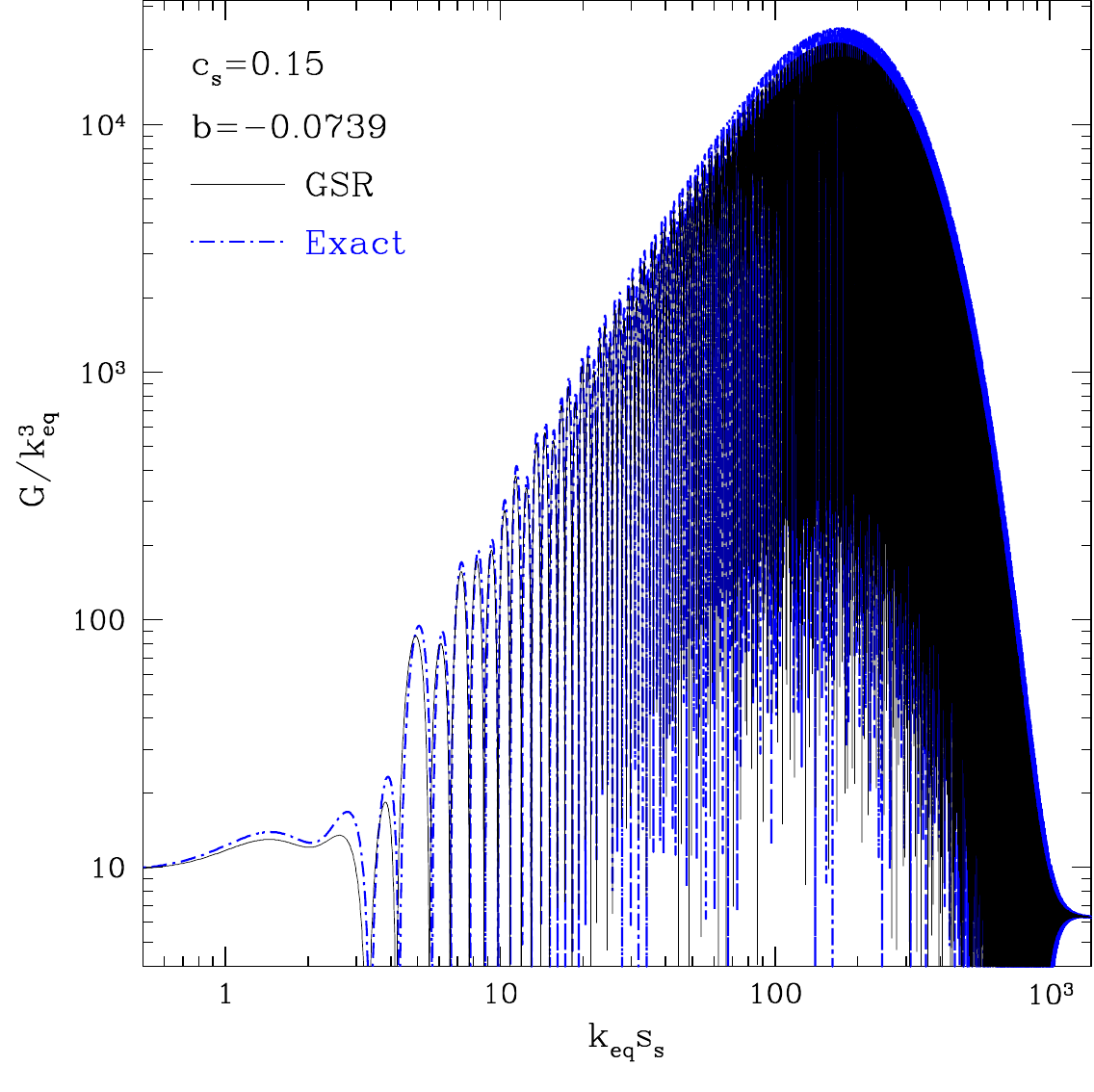, width=3in}
\psfig{file=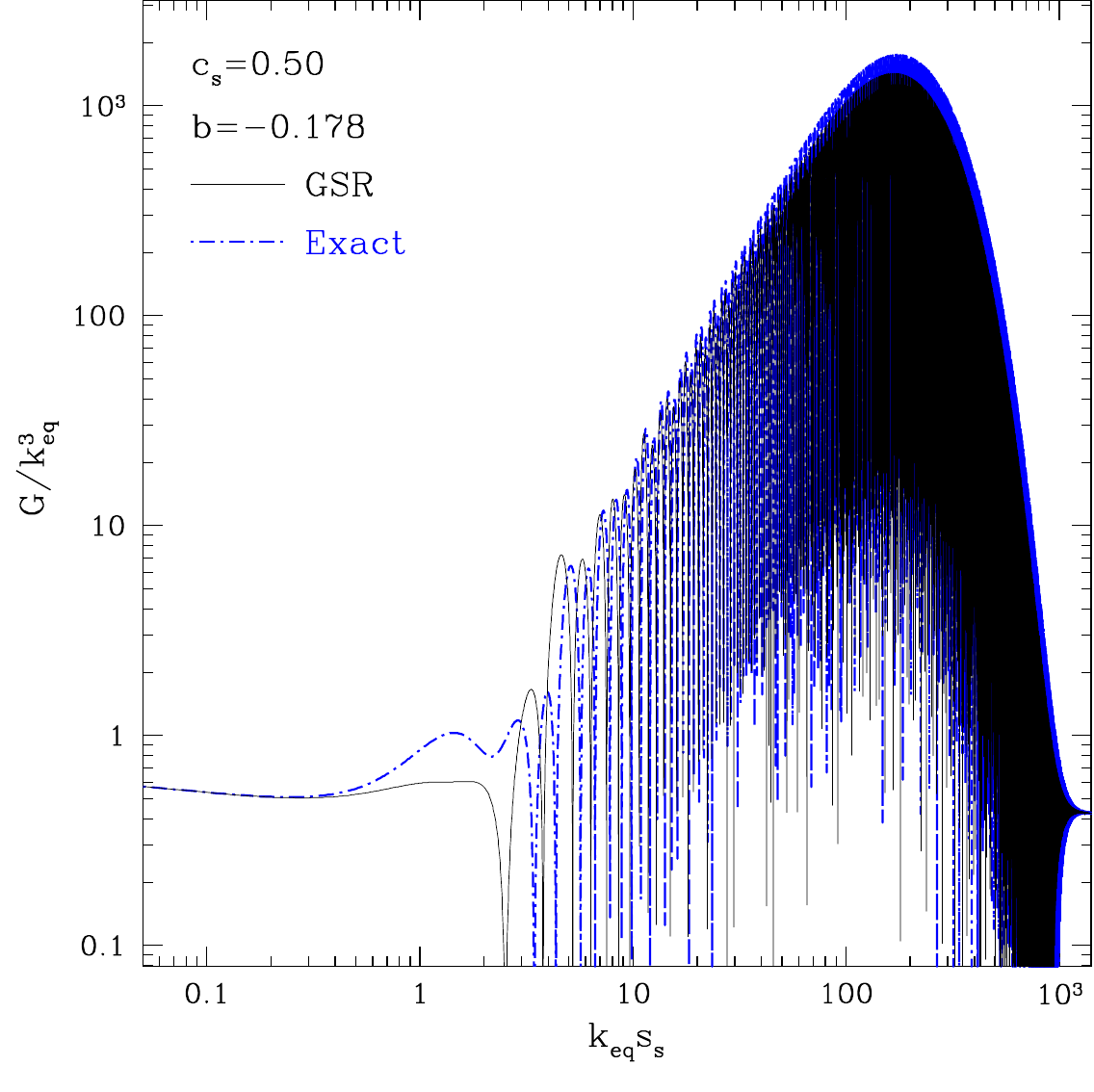, width=3in}
\caption{Exact vs.\ GSR approximation for equilateral triangles. Shown are a
 small $b=-0.005$ amplitude warp step (top) and the large amplitude step that matches the  \emph{Planck} feature (bottom) for a low
 $c_s=0.15$ (left) and high $c_{s}=0.5$   (right) sound speed.   For the small amplitude step the
 GSR approximation is accurate in amplitude at the several percent level.   For
 the large amplitude step the amplitude accuracy is $\sim 20\%$ or better at peak.  Other parameters have been set to reproduce the Planck best fit, in particular $x_d=87.4$.
}
\label{fig:GSRvExact}
\end{figure*}

Fig.~\ref{fig:GSRvExact} (top) shows the result for a small amplitude feature
with $b=-0.005$  for two different values of $c_s=0.1, 0.5$.    The GSR approximation captures the amplitude of the features to a few percent  for both cases.  
In particular the amplitude increases as $(k_{\rm eq} s_s)^2$ until it reaches a 
peak at $k_{\rm eq} s_s \approx 2x_d$ before damping away.
As we shall see below, varying the sound speed changes the relative weight of the different operators and so
the two different cases demonstrate that the relevant terms have been individually
calculated correctly.
  
In Fig.~\ref{fig:GSRvExact} (bottom), we test the larger amplitude feature preferred
by the  \emph{Planck} data.    Given that the same fractional change in the warp or sound speed causes
a smaller effect as $c_s \rightarrow 1$, the amplitude of the step $b$ required to match the
data increases with $c_s$.   On the other hand the accuracy of the GSR approximation 
depends directly on $b$ and so errors increase with $b$ and $c_s$.  

For $c_s = 0.15$, the  \emph{Planck} parameters predict an extremely large equilateral bispectrum
at peak.   We illustrate this value since the maximum violates even a weak
criteria for  the validity of perturbation theory
\begin{equation}
\frac{\cal G}{k_{\rm eq}^3} \Delta_\curv \lesssim 1.
\label{eqn:PT}
\end{equation}
Beyond this point, the curvature field is strongly non-Gaussian and loop corrections likely invalidate the calculation. Note that this sets a firm lower bound on $c_s$ for models that seek to explain the  \emph{Planck} feature with steps in the warp.

For the case of $c_s=0.5$, the full change across the step approaches order unity, specifically  $| 2 b | \approx 0.36$ 
and the
GSR approximation at the peak of the bispectrum holds to $\sim 20\%$.    For larger
values of $c_s$ it is impossible to explain the feature without a very large step in 
the warp.    Note a step in the potential does not suffer this problem and they
remain viable explanations even as $c_s \rightarrow 1$.   Likewise for a potential
step in canonical single field inflation with the same parameters that fit the  \emph{Planck} power spectrum
$|{\cal G}/k_{\rm eq}^3|_{\rm max} \approx 766$.  The amplitude of the equilateral bispectrum
clearly distinguishes the two scenarios for $c_s \lesssim 0.5$.

\begin{figure}[t]
\psfig{file=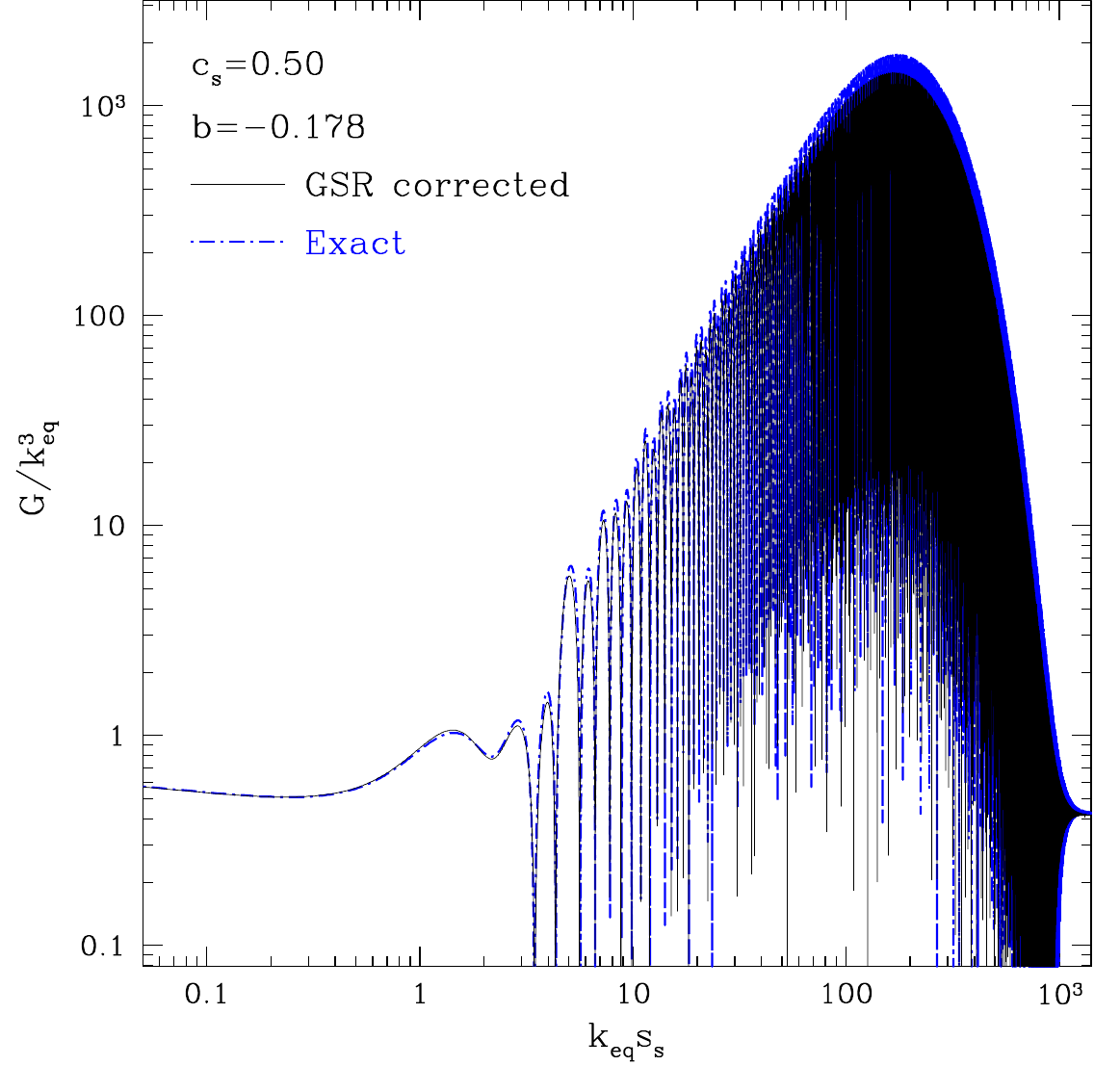, width=3in}
\caption{Modified GSR source correction for the highest amplitude $b$  case of
Fig.~\ref{fig:GSRvExact}.   By changing the $i=2$ modefunction correction source
to a total derivative of a slow-roll suppressed quantity through Eq.~(\ref{eqn:modsource}), the errors at $k s_s \sim 1$ decrease. }
\label{fig:modsource}
\end{figure}

The errors in the 
GSR approximation at high $b$ are mainly due to the $i=2$ operator.     
Note that in obtaining
the modefunction correction source to this term in Eq.~(\ref{eqn:S2m}), we have assumed that $S_2$ in
Eq.~(\ref{eqn:S2}) is constant whereas it is a function of $c_s$.  Corrections to this
approximation from the sound speed step are expected to be important as the amplitude
of the step increases.   

In fact the errors in the approximation around $k_{\rm eq} s_s \sim 1$ can be 
directly attributed to this problem.    If $x_d \gg 1$ subleading terms in the approximation
begin to dominate here and cause order unity errors.   Fortunately, this occurs only
in the region where the bispectrum is too small to be observed due to cosmic variance.
We can nonetheless correct for this problem by modifying the source $S_{2m}$ for $m\ge 3$
from the $S_{2m}'=g S_2$ form of Eq.~(\ref{eqn:S2m}).  The problem is that this form
is not a total derivative of a slow-roll suppressed source and so for large amplitude
features it does not integrate back to a slow-roll suppressed $S_{2m}$ after the inflaton
has transited the feature.     Analogous effects in the $i=0$ operator are fixed by
self consistently expanding to the next order in the GSR approximation
\cite{Adshead:2011bw}.   
Since this would involve nested integrals, we can fix the problem by simply replacing
the source with a form that is identical in the slow-roll and small feature limits but
which carries the total derivative structure
\begin{equation}
S_{2m} \rightarrow \frac{3}{2} [G(\ln s)-\bar G]S_2 .
\label{eqn:modsource}
\end{equation}
To ensure that the source is slow-roll suppressed after the feature we set the
constant $\bar G = G(\ln s_s; b=0)$.    In Fig.~\ref{fig:modsource} we show that this change corrects the $k s_s \sim 1$ problem.   On the other hand, this problem
appears in a non-observable part of the spectrum for this model and also
disappears if $x_d  \lesssim 25$, where one might otherwise think $k s_s \sim 1$ effects
are important, and so we do not consider it further.

\begin{figure}[t]
\psfig{file=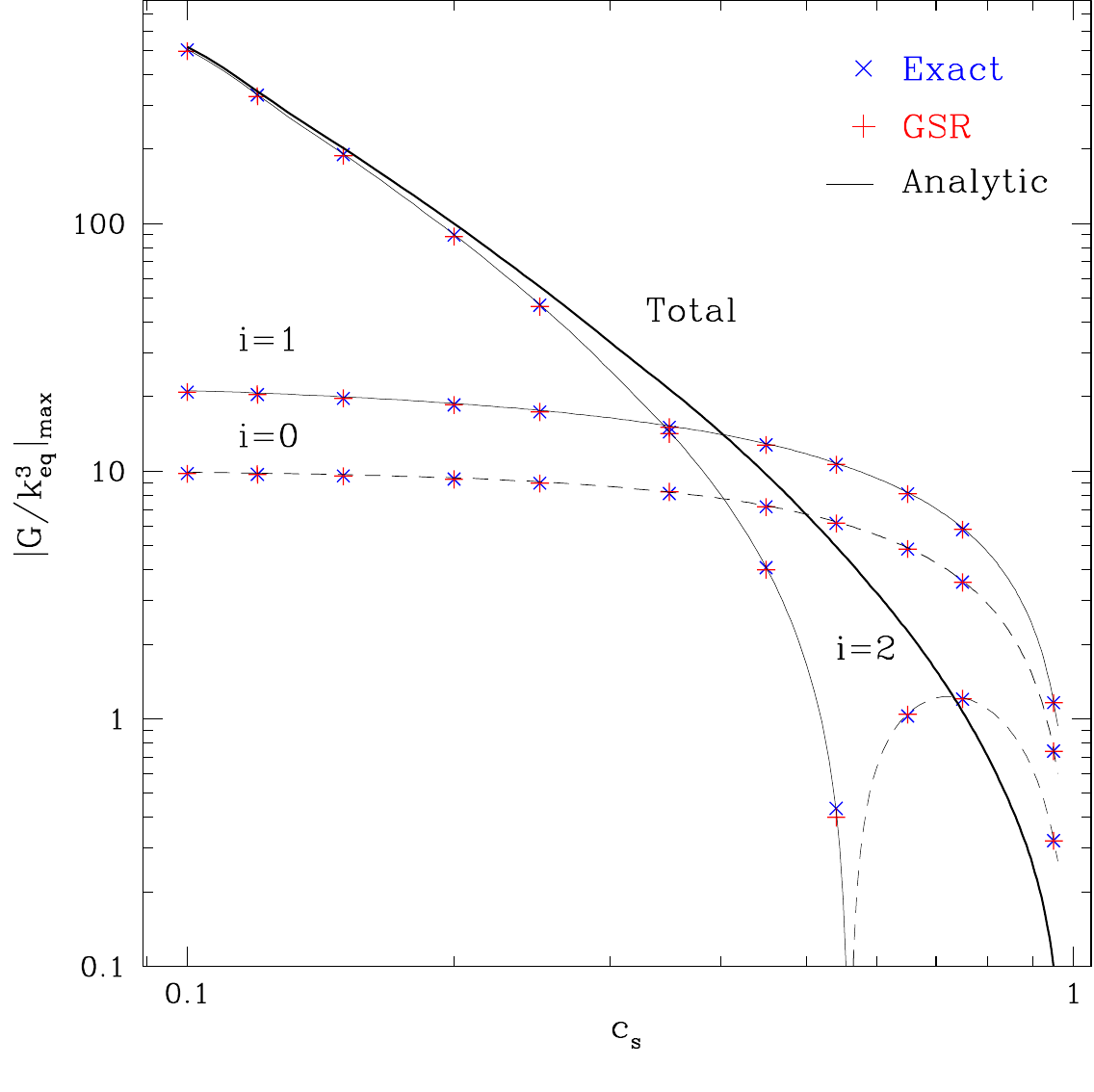, width=3.0in}
\caption{Comparison between exact, GSR and leading order analytic solution predictions for the maximum amplitude of $|{\cal G}/k_{\text{eq}}^3|$. The step parameters are $b=-0.005$,  $x_d = 30.2$. Dashed lines indicate where the individual contributions have the opposite sign to total.  For $c_s\ll 1$ the $i=2$ term
dominates whereas for $c_s \rightarrow 1$ all terms are comparable.}
\label{fig:analytic}
\end{figure}

\subsection{Analytic Scaling}
\label{sec:DBIanalytic}

Given  the sensitivity of the maximum equilateral bispectrum amplitude to the choice of parameters and the large current observational uncertainties on their values from the
 \emph{Planck} data, it is useful to have analytic scalings for the amplitude and form around maximum.

In the sharp-feature limit, the dominant contributions to the bispectrum arise from the terms
with the highest number of temporal derivatives and the windows with the steepest scalings with
$x$, i.e.\ those that involve $x \sin x$ in our source convention. 
In principle we can also keep subleading terms in the analytic expression for the bispectrum. The rather involved results are neither illuminating nor necessary near the peak of the bispectrum as long as $x_d\gg 1$
 and so we omit them here.

It is straightforward to derive an analytic expression for their contribution in the $b\rightarrow 0$ limit.   Integrating once by parts
\begin{align}\label{eqn:smallstepdom}
I_{00} \approx &\int \frac{ds}{s} \frac{\eta_H-\sigma_1/2}{f} X(K s), \nonumber\\
I_{10} \approx  &- \int \frac{ds}{s} \frac{\sigma_1}{f} X(K s), \nonumber\\
I_{20} \approx & \int \frac{ds}{s}  \left[ -\frac{2}{c_s^2}\frac{\sigma_1}{f} -\left( 1 - \frac{1}{c_s^2} \right)\frac{\eta_H+3\sigma_1/2}{f} \right] X(K s) ,\nonumber\\
I_{23} \approx & \int \frac{ds}{s} \left( 1- \frac{1}{c_s^2}\right) \frac{\eta_H + \sigma_1/2}{f}X(K s) ,
\end{align}
where $X(x) = x^2 \cos x$.   Combining these terms, we can approximate the bispectrum as
\begin{eqnarray}
\frac{\cal G}{k_1 k_2 k_3} &\approx& -\frac{\Delta_\curv(k_1) \Delta_\curv(k_2) \Delta_\curv(k_3)}{4 A_s^2} 
\\
&& \times \int \frac{ds}{s}  \left[  \frac{\sum_i k_i^2}{K^2}  \frac{\sigma_1}{c_s^2 f}+\frac{\eta_H}{c_s^2 f} \right] X(K s). \nonumber
\end{eqnarray}

The slow-roll parameters $\sigma_1$ and $\eta_H$ may be approximated as \cite{Miranda:2012rm}
\begin{align}
	\sigma_1  &\approx (1-c_s)  b \frac{ d F}{d\ln a} , \nonumber\\
	 \eta_H &\approx  -\frac{c_s}{2} \frac{1- c_s}{1+ c_s} b \frac{ d F}{d\ln a},
\end{align}
where recall that $F$ is a step-like function in $\ln a$.  As $d \rightarrow 0$,  $d F/ d\ln a$ can be approximated as a delta function with normalization set by $\int d\ln s( dF/d\ln a) \approx -\int d F \approx 2$. 
For finite step width $d$ in field space, the inflaton traverses the step in 
$\Delta s/s_s \approx  |d\ln s/d\phi | d$.   The window functions $X$ oscillates on a time scale $\Delta s=1/k$.   Thus the integral
is damped for $k s_s > \phi'/d$ and the same characteristic damping scale as in the power spectrum
$x_d = \phi'/\pi d$  (see Eq.~\ref{eqn:xd}) appears here.   For the tanh step, the integral can be approximated following \cite{Adshead:2011jq}
\begin{eqnarray}\label{eqn:analyticbi}
\frac{\cal G}{k_1 k_2 k_3} &\approx& -\frac{\Delta_\curv(k_1) \Delta_\curv(k_2) \Delta_\curv(k_3)}{4 A_s^2}
{\cal D}\left( \frac{K s}{ 2 x_d}\right) \frac{ X(K s)}{c_s^2 f} 
\nonumber\\
&& \times\left[   { 2(1-c_s) }\frac{\sum_i k_i^2}{K^2} -c_s \frac{1- c_s}{1+ c_s}    \right] b \Bigg|_{s_s},
\end{eqnarray}
where
the damping function is
\begin{equation}
{\cal D}(y) =  \frac{y}{\sinh y}.
\end{equation}
In particular, for equilateral triangles, the bispectrum reaches a maximum of
\begin{equation}
\bigg| \frac{\cal G}{k_{\rm eq}^3} \bigg|_{\rm max}\approx 10.77 \frac{(2-c_s)(1-c_s)}{{12}(1+ c_s)}  \frac{|b|}{c_s^2}  x_d^2 (2 x_d)^{3(n_s-1)/2}
\end{equation}
at a scale of approximately $k s_s \approx 2 x_d$.

In Fig.~\ref{fig:analytic} we compare this analytic expectation with the GSR and exact
calculations at the peak of the equilateral bispectrum.  For the analytic curves, we plot the analog of Eq.\ (\ref{eqn:analyticbi}) for the terms in Eq.\ (\ref{eqn:smallstepdom}).  For a small amplitude
step, the agreement between all three is excellent.  The $i=2$ term  {($I_{20}+I_{23}$)} dominates at $c_s \ll 1$
and exhibits the $c_s^{-2}$ enhancement of the equilateral bispectrum over power spectrum
effects which scale as $2(1-c_s)b/(1+c_s)$ through Eq.~(\ref{eqn:poweramp}) \cite{Miranda:2012rm}. 
As $c_s \rightarrow 1$ all terms are comparable and in fact
partially cancel each other.  While the total is suppressed in this limit, it scales in the
same way with $(1-c_s)$ as the power spectrum.    

Fig.~\ref{fig:GSRvExact} verifies that the scaling remains qualitatively correct for the
 \emph{Planck} amplitude step.
The rough criteria for the validity of perturbation theory given in Eq.~(\ref{eqn:PT}) becomes
\begin{equation}
10.77 \frac{(2-c_s)(1-c_s)}{{12}(1+ c_s)}  \frac{|b|}{c_s^2}  x_d^2 (2 x_d)^{3(n_s-1)/2}
 \Delta_\curv \lesssim 1.
\end{equation}
Thus sound speeds smaller than the $c_s=0.15$ example in the previous section 
would become allowed if  $\sqrt{|b|} x_d$ dropped significantly.

We note that the result we find for the bispectrum here is 
different from the estimate in Ref.~\cite{Bean:2008na}. The key difference is that Ref.~\cite{Bean:2008na} approximated the enhancement to the bispectrum as arising for modes near horizon crossing due to the slow-roll parameter $\eta_H$ becoming large, and therefore the modes interacting more strongly near where they were freezing out. {While this is approximately true for models with features in the warp or potential that are crossed on a timescale on the order of an efolding, in this work we have demonstrated that} the key effect of the sharp feature in the background is that correlations are frozen at the time of the feature, $s_s$, rather than near horizon crossing, $s \sim 1/k$. This means that correlations between curvature perturbations with momentum $k  \gg 1/s_s $ are imprinted well before horizon crossing where the amplitude of the fluctuations is much larger and is oscillating with varying $k$. This results in a strongly scale dependent and oscillatory bispectrum that is
only cut off at the damping scale derived above.{ We note that improved estimates of the scaling behavior due to a sharp step were presented in Ref.\ \cite{Chen:2011zf}.}  In addition, the relative contribution of
$\eta_H$ and $\sigma_1$ terms as a function of width $d$ was misestimated in Ref.~\cite{Bean:2008na}.

\section{Discussion}\label{sec:conclusions}

In this work we have studied the bispectrum that arises in general models of single field inflation beyond the slow-roll expansion. 
Our integral approach allows the expansion history and inflaton sound speed to be 
arbitrary functions of time and encompasses all terms in 
 the effective field theory of inflation aside from those involving the extrinsic curvature and Galileon interactions. This form allows for a fast computation of all bispectrum configurations
 from a handful of one dimensional integrals and 
 should facilitate efficient comparison of these models with data.
 
 The key assumption of our approach is that the expansion remains nearly de Sitter with $\epsilon_H \ll 1$ during inflation, while $\epsilon_H$ is allowed to have arbitrary variation. In particular, we explicitly drop operators that contribute to the bispectrum as $\epsilon_H^2$ and approximate
 the evaluation of the rest assuming that the curvature modefunctions remain
 perturbatively close to their de Sitter forms.

Motivated by recent power spectrum analyses of  \emph{WMAP} \cite{Adshead:2011jq} and  \emph{Planck} \cite{Ade:2013rta} data, we take as an illustrative example  a sharp step in the warp-brane tension in the context of DBI inflation \cite{Bean:2008na}. We show that our integral approximations are excellent for small amplitude steps in the warp, while for 40\%  steps remain accurate  to $20\%$.  Further, we demonstrate that the  consistency relation between the squeezed limit of the bispectrum and the power spectrum spectral index is respected in our approximations. 

A step in the sound speed due to a step in the warped-brane tension large and sharp enough to explain the high frequency oscillations in the  \emph{Planck} power spectrum would also generate a very large bispectrum peaking in the equilateral limit. For low sound speeds, $c_s$, these bispectra are large enough to violate even weak criteria for the validity of perturbation theory and effectively put a lower bound of $c_s > 0.15$.  Furthermore, as the sound speed increases, the step in the warp must become larger and larger in order to produce the same power spectrum features and ceases to remain viable as $c_{s}\to 1$. In particular, for $c_s > 0.5$, one requires an order unity step in the warp to produce the favored 10\% oscillations in the power spectrum. We have also demonstrated that otherwise degenerate scenarios  that produce oscillations in the power spectrum -- steps in the warp or potential -- are distinguished by their various bispectra. For $c_s < 0.5$, the scenarios are clearly distinguished by the amplitude of their equilateral bispectra.

Our technique also applies to the cases where the slow-roll approximation remains valid, but where slow-roll corrections may be large \cite{Burrage:2011hd}.   We provide a complete and
compact first-order expression for all bispectrum configurations.   In Appendix \ref{app:litcomp} we demonstrate the consistency of our results with previous work \cite{Burrage:2011hd} for
several limiting cases.
Given the observed value of the tilt, we show that large slow-roll corrections only exist in two cases: where one or more of the slow-roll parameters $\{\epsilon_H,\eta_H,\sigma_1\}$ is anomalously larger than the tilt $n_s-1$ or when the sound speed 
{$c_s \gtrsim 0.8$.}

Our work is somewhat orthogonal to the work of Ref.\ \cite{Noller:2011hd} who evaluated the bispectrum for a class of $P(X, \phi)$ theories allowing $\epsilon_H \sim 1$ and $\sigma_1 \sim 1$ while higher slow-roll parameters, $\sigma_2$, $\eta_H$, etc., were assumed small. Further, Ref.\ \cite{Ribeiro:2012ar} considered the bispectrum in Horndeski theories, relaxing the slow-roll assumption, again allowing for $\epsilon_H \sim 1$ and $\sigma_1 \sim 1$ in such a way as to preserve a scale invariant spectrum. However, in the region where our analyses overlap, one can demonstrate our results are equivalent to those of \cite{Noller:2011hd} (see Appendix \ref{app:litcomp} and Refs.\ \cite{Burrage:2011hd, Ribeiro:2012ar}).

The expressions presented here are ideally suited for use in concert with a fast estimator for the angular bispectrum in CMB. In particular this form should facilitate  searches for the non-Gaussian counterparts to features in the power spectrum data which could confirm their
primordial origin.

\acknowledgements

We thank Mark Wyman for useful discussions.
PA and WH
were supported by the Kavli Institute for Cosmological
Physics at the University of Chicago through grants NSF
PHY-1125897 and an endowment
from the Kavli Foundation and its founder Fred Kavli.
VM and WH were additionally supported by U.S.\ Dept.\ of
Energy contract DE-FG02-90ER-40560,  WH by the David and Lucile
Packard Foundation and VM by the Brazilian Research Agency
CAPES Foundation and by U.S. Fulbright Organization.
\wh{We thank Sam Passaglia for pointing out a typo in $T_{40}$.}


\appendix

\section{Completing the Triangle}\label{app:additionalterms}

In the main text, we deferred consideration of two operators that were not important in the
DBI step example.   In this Appendix we complete the GSR integral formulation for the bispectrum of general
single field inflation.

\subsection{$i=3$: $\dot \curv^3$}
\label{sec:Rdot3term}

Beyond DBI models, general single field inflation models can also have $c_s^{-2}$ enhanced
bispectra through the $\dot \curv^3$ term in the cubic action, Eq.~(\ref{3action}).  In the $P(X,\phi)$ model 
it is given by Eq.~(\ref{eqn:XiP}).  In the general effective theory of inflation it is associated
with the $M_3$ mass scale defined in Ref.~\cite{Cheung:2007st} or equivalently the 
$c_3 = -M_3^4/M_2^4$ coefficient of Ref.~\cite{Cheung:2007sv}. For a $P(X, \phi)$ theory, in our notation
\begin{align}
M_n^4(t) =&  \left(-{X}\)^n\frac{\partial^n P}{\partial {X}^n}\Big|_{X=\bar X(t)}, \nonumber\\
 \bar{X}(t) = &
\frac{1}{2} \left( \frac{d \bar\phi}{dt} \right)^2,
\end{align}
where $\bar\phi$ is the homogeneous background scalar field. With these conventions, $M_2^4= \Sigma (1-c_s^2)/2$ and
\begin{align}
\frac{M_3^4}{M_2^4} =& 3 \left( \frac{1}{2} -\frac{1}{1-c_s^2} \frac{\lambda}{\Sigma} \right) \nonumber\\
=& \frac{3}{2} \left( 1 -\frac{1}{c_s^2} \right), \quad {\rm DBI}
\end{align}
(cf.\  \cite{Achucarro:2012fd} Eq.~5.12).   For effective theories that parametrize turning trajectories in
 multifield models where the heavy degrees of freedom have been integrated out this scale
 becomes a specific function of the time-varying sound speed $c_s$ \cite{Achucarro:2012fd}.
 In summary,
\begin{eqnarray}
\Xi =
\begin{cases}
  0 , & \quad {\rm DBI}\\
  \dfrac{1}{c_s^2}-1 - 2 \dfrac{\lambda}{\Sigma} , & \quad P(X,\phi) \\
 \dfrac{ (1- c_s^2)^2}{2 c_s^2}, & \quad {\rm turn} \\
 \dfrac{ (1-c_s^2)^2}{c_s^2}  - \dfrac{2}{3} (1- c_s^2) c_3. & \quad {\rm EFT}
 \end{cases}
 \end{eqnarray}
 It is interesting to note that 
 \begin{equation}
 P(X,\phi) = T(\phi)\left[ 1- \frac{T(\phi)}{T(\phi)-X} \right]  - V(\phi)
 \end{equation}
 would mimic the turning trajectory case at the effective field theory level.

 Inserting the $\dot\curv^3$ operator into the general expression Eq.~(\ref{eqn:bispec}),
 we obtain
\begin{eqnarray}
\frac{\cal G}{k_1 k_2 k_3} &\supset&  \frac{3}{4 A_s^2} {\cal R} \Bigg(
i \left[ \prod_{i=1}^3 \frac{ x_{i*} y_{i*}}{f_*} \right] \int_{s_*}^\infty \frac{ds}{s} S_{3}(\ln s) \nonumber\\
&&\times
\prod_{i=1}^3\left[\frac{f}{x_i} \left( \frac{ x_i y_i^*}{f} \right)'  \right]  \Bigg) ,
\label{eqn:Rdot3}
\end{eqnarray}
where
\begin{equation}
S_3(\ln s) = {-}\frac{c_s}{a H s}\frac{\Xi}{f} .
\end{equation}

As with the terms in  the main text, an exact evaluation of the $\dot \curv^3$  bispectrum contribution involves
first solving numerically for the modefunction $y$ through the exact equation of 
motion Eq.~(\ref{eqn:yeqn}).   Note that as $x \rightarrow \infty$, $y \rightarrow e^{i x}$
and so the term in the second bracket tends to diverge as 
$s^3 e^{-3 i Ks}$
making it challenging to evaluate numerically.   In practice when evaluating the exact bispectrum contributions, we regulate such expressions with an artificial damping factor at a sufficiently large $s$ that the model is in the slow-roll
regime.

The GSR approximation can be constructed to avoid such problems.  There are
two types of terms in general: those that involve replacing $y \rightarrow y_0$ and taking $f \approx $ const. in
Eq.~(\ref{eqn:Rdot3}) and those that involve the first-order modefunction correction
from Eq.~(\ref{eqn:firstordermode}) or $f'/f$.   The latter is required since $\Xi$ contributes
at zeroth order in the slow-roll expansion.  

For the former case 
\begin{eqnarray}
\frac{\cal G}{k_1 k_2 k_3} &\supset&  \frac{\Delta_\curv(k_1) \Delta_\curv(k_2) \Delta_\curv(k_3)}{A_s^2}\frac{3}{4}\frac{k_1 k_2 k_3}{K^3}   
\nonumber\\
&&\times \int_{s_*}^\infty \frac{ds}{s} S_{3}(\ln s) (K s)^3 \sin (K s),
\end{eqnarray}
where we have replaced $1/f_* \rightarrow \Delta_\curv$ as appropriate for zeroth order
modefunction expressions (see below). 
While compact in form, this expression is again numerically difficult to evaluate at 
$K s \rightarrow \infty$.  Instead, we first integrate this expression twice by parts
and bring the result to the standard form of Eq.~(\ref{eqn:GSRBi})
\begin{equation}
\frac{\cal G}{k_1 k_2 k_3}  \supset \frac{\Delta_\curv(k_1) \Delta_\curv(k_2) \Delta_\curv(k_3)}{4 A_s^2} \sum_{j=0}^2 T_{3j} I_{3j}(K) ,
\end{equation}
with the sources
\begin{eqnarray}
S_{30}&=&S_{31}= S_3' ,\nonumber\\
S_{32}&=& S_3,
\end{eqnarray}
windows
\begin{eqnarray}
W_{30}(x) &=& x \sin x ,\nonumber\\
W_{31}(x) &=& W_{32}(x) = \cos x,
\end{eqnarray}
and triangle weights
\begin{eqnarray}
T_{30} = T_3, \quad T_{31} = 3 T_3, \quad T_{32}=2 T_3,
\label{eqn:T3nm}
\end{eqnarray}
where
\begin{eqnarray}
T_3 = -\frac{3k_1 k_2 k_3}{K^3}.
\end{eqnarray}

The modefunction correction terms involve both corrections to the external modefunctions evaluated at $s_*$ and nested integrals involving corrections inside the original integral in
 Eq.~(\ref{eqn:Rdot3}).   The former type is approximated by the replacement of
 $1/f_* \rightarrow \Delta_\curv$ above.  Note that the out of phase type contribution discussed in Ref.~\cite{Adshead:2012xz} vanish in this case since the $(K s)^3 \cos(K s)$ window integrates to zero.  The latter type involves integrals over $g(\ln s)$ in
 Eq.~(\ref{eqn:firstordermode}).
   Part of the $\curv$ modefunction correction term involves derivatives
 acting on $f$ in  Eq.~(\ref{eqn:Rdot3}).   This can also be brought into nested form by
\begin{equation}
\frac{f'}{f} \approx s^3 \int d\ln s \frac{g}{s^3}.
\end{equation}
The resulting terms can be simplified by integration by parts given that to leading order
$S_3$ can be taken to be constant here.   The result is that the new source 
becomes
\begin{equation}\label{eqn:gS3}
g S_3 = S_{33}'=S_{34}'=S_{35}' ,
\end{equation}
with now all terms combined into the form
\begin{eqnarray}
\frac{\cal G}{k_1 k_2 k_3}  &\supset&  \frac{\Delta_\curv(k_1) \Delta_\curv(k_2) \Delta_\curv(k_3)}{4 A_s^2} \Big\{  \sum_{j=0}^5 T_{3j} I_{3j}(K)  \nonumber\\ \qquad && + [
T_{3B} I_{35}( 2 k_3) +{\rm perm.}] \Big\}.
\end{eqnarray}
Here the additional windows are
\begin{eqnarray}
W_{33}(x) &=& x \sin x + \cos x, \nonumber\\
W_{34}(x) &=& 2\frac{\sin x}{x} - \cos x ,\nonumber\\
W_{35}(x) &=& 12\left(  \frac{\sin x}{x^3} - \frac{\cos x}{x^2} -\frac{\sin x}{ 4 x}\right),
\end{eqnarray}
and triangle weights
\begin{eqnarray}
T_{33} &=& \frac{3 k_1 k_2 k_3}{K^2 (K- 2 k_3)} +{\rm perm.},\nonumber\\
T_{34} &=& - \frac{3 k_1 k_2 k_3}{K^2 (K- 2 k_3)^2} \left[ 7(k_1+k_2)-3 k_3 \right] +{\rm perm.},
\nonumber\\
T_{35} &=& \frac{4 k_1 k_2 k_3}{K^2 (K-2 k_3)^3} \left[ 5 (k_1+k_2)^2 - 5(k_1+k_2) k_3 + 2 k_3^2 \right]
\nonumber\\ &&
+ {\rm perm.} \nonumber\\
T_{3B} &=& -\frac{2  k_1 k_2 k_3}{(K-2 k_3)^3}.
\label{eqn:T3m}
\end{eqnarray}
The final $T_{3B}$ term comes from the boundary term from unnesting the integrals
through integration by parts.
The boundary terms in $I_{ij}$ from the modefunction terms cancel since
\begin{equation}
\sum_{j=3}^5 T_{3j} + (\wh{T_{3B}} + {\rm perm.}) = 0.
\label{eqn:Tcancel3}
\end{equation}

The $s_*$ boundary term in $I_{32}$ gives zeroth order slow-roll result for the $P(X,\phi)$ model as
\begin{eqnarray}
 \frac{\cal G}{k_1 k_2 k_3} &\approx & \frac{3}{2}  \frac{k_1 k_2 k_3}{K^3} \left(  \frac{1}{c_s^2} -1- 2\frac{ \lambda}{\Sigma} \right),
 \end{eqnarray}
 which reproduces a well-known result  \cite{Seery:2005wm,Chen:2006nt}.

\subsection{$i=4$: $ \dot\R  \partial_i\R \partial_i \chi  $}
\label{sec:chiterm}

Finally, for completeness we consider the (non-local) $ \dot \curv \partial_i \curv   \partial_i \chi $ operator in
Eq.~(\ref{3action}) that was dropped in the main paper.    Its contribution can be cast as a source
\begin{align}
	S_{40} = S_{41} = \frac{\epsilon_H}{ c_s^2 f},
\end{align}
with windows
\begin{eqnarray}
	W_{40}(x) & = & x \sin x, \nonumber \\
	W_{41}(x) & = & \cos x,
\end{eqnarray}
and triangle weights 
\wh{
\begin{eqnarray}
T_{40} &=& T_{41} - \frac{(K-2 k_1)(K-2 k_2)(K-2 k_3)}{k_1 k_2 k_3}, \label{eqn:T4} \\
T_{41} &=& \frac{1}{k_1 k_2 k_3} \bigg(
	\frac{K}{2} \sum_i k_i^2 - \frac{3}{2} \sum_i k_i^3 + \frac{2}{ K^2} \sum_{i\ne j} k_i^2 k_j^3 \bigg) .
	\nonumber
\end{eqnarray}
}
Note that in the squeezed limit of $k_1=k_S \ll k_L=k_2\approx k_3$, both weights scale as
$k_S/k_L$ and do not contribute.   Hence they are not required for the establishment of the slow-roll
consistency relation.    On the other hand $T_{41}$ does contribute to the $\epsilon_H$ terms in the slow-roll bispectrum for other triangles.

\section{Slow-Roll Expansion}
\label{app:sr}

With the completion of the GSR expression in \S \ref{app:additionalterms}, we can
expand the full result to first order in the slow-roll parameters
$\epsilon_H$, $\eta_H$ and $\sigma_1$ assuming they are nearly constant.  This expansion trivially reproduces the usual slow-roll approximation for the
$i=0,1,4$ operators and improves their accuracy by defining the respective epochs of freezeout near horizon crossing.   On the other hand, the $i=2,3$ operators involve sources that are zeroth order in
slow-roll parameters and thus complicate the slow-roll expansion \cite{Chen:2006nt,Burrage:2011hd}.

\subsection{GSR Derivation}

In the slow-roll limit, all of the sources in the GSR integrals can be taken to be slowly varying
\begin{equation}
S_{ij}(\ln s) \approx S_{ij}(\ln s_*) + S_{ij}'  (\ln s - \ln s_*),
\end{equation}
where $S_{ij}'$ is taken to be a constant.  Thus the integrals reduce to 
\begin{equation}
I_{ij} = S_{ij} (\ln s_*)W_{ij}(x_*)  +S_{ij}'  \int_{x_*}^\infty \frac{d x}{x} W_{ij}(x),
 \end{equation}
 where $x_*=Ks_*$.
For the windows where $\lim_{x\rightarrow 0} W_{ij} \ne 0$, the integral can be
re-expressed as the evaluation of the source $S_{ij}$ 
\begin{equation}
I_{ij} = S_{ij}(\ln s_f) W_{ij}(x_*)
\end{equation}
at the freezeout epoch 
$s_f = x_f/K$ where
  \begin{equation}
 \ln x_f - \ln x_* = \int_{x_*}^\infty \frac{d x}{x} \frac{W_{ij}(x)}{W_{ij}(x_*)}.
 \end{equation}
 In Tab.~\ref{tab:freezeout}, we give $x_f$ for the various GSR windows.

\begin{table}[tb]
\begin{center}
\begin{tabular}{ c   c }
$W_{ij}(x) $  &  $x_f$ \\
\hline
$\cos x$  & $e^{-\gamma_E}$ \\
$\dfrac{\sin x}{x}$ & $e^{1-\gamma_E }$ \\
$W(x/2)$ & $e^{7/3-\gamma_E}$ \\
$W(x)$ & $e^{7/3-\gamma_E} /2$ \\
$W_n(x)$ & $e^{2 -\gamma_E}/2$ \\
\end{tabular}
\end{center}
\caption{Freezeout epoch $x_f$ for the various GSR windows where $\gamma_E \approx
0.5772$ is the Euler-Mascheroni constant, $W$ is the power spectrum window of
Eq.~(\ref{eqn:powerwindow}) and $W_n$ is the tilt window of Eq.~(\ref{eqn:tiltwindow}).}
\label{tab:freezeout}
\end{table}%

The GSR integrals can be further simplified by approximating the sources to first
order in the slow-roll parameters through 
\begin{align}\label{eqn:srcsrexp}
\frac{a H s}{c_s}  &= -(\ln a)' \approx  1+ \epsilon_H + \sigma_1 ,\nonumber\\
\frac{( a H s)'}{a Hs} &\approx  - \sigma_1,  \nonumber\\
G' & \approx \frac{2}{3}g  \approx - 2 \frac{f'}{f} \approx  1-n_s\nonumber\\
&\approx  ( 4\epsilon_H - 2\eta_H + \sigma_1) .
\end{align}
With these approximations, the bispectrum to first order in the slow-roll parameters
is given by 
\begin{align}\label{eqn:srfull}
\frac{ {\cal G}}{k_1 k_2 k_3} \approx&\frac{1-n_s}{8 } \bigg[ T_{01}+ T_{02}    
+3\left( \frac{1}{c_s^2} -1\right)  T_{2n} - 3\Xi T_{3n} \bigg]
\nonumber\\
& 
+ \frac{\epsilon_H + \sigma_1}{4 } ( T_{11}+T_{12})   -\left( \frac{c_s}{ a H s}\frac{\Xi}{f} \right)' \frac{f}{4} T_{31} 
  \nonumber\\&
 +\frac{\epsilon_H}{4 c_s^2} T_{41} 
 +  \frac{(1-c_s^2)(1-n_s) + 4\sigma_1}{8 c_s^2 } T_{21}
   \nonumber\\
&  
+ \frac{ \Delta_\curv(k_1) \Delta_\curv(k_2) \Delta_\curv(k_3) }{
4 A_s^2 }\frac{1}{f}  \frac{c_s}{a H s}   
\nonumber\\
& \times \bigg[  \left( \frac{1}{c_s^2} -1\right)  T_{22}-\Xi T_{32}
\bigg],
\end{align}
where the final source is evaluated at $K s = e^{-\gamma_E}$ and we have assumed the slow-roll parameters $\epsilon_H, \eta_H$, and $\sigma_1$ are constant in dropping  terms.  
The accuracy can
be further improved by evaluating the slow-roll parameters at their respective freezeout
epochs rather than taking them to be constant. 
In fact the modefunction correction terms contribute through these freezeout relations as
\begin{eqnarray}
T_{2n}  &=&  T_{23} + 2 T_{25} + \frac{7}{3} T_{26} 
\nonumber\\
&&+ \left[ T_{2B}\left( \ln \frac{K}{2 k_3}+\frac{7}{3}\right) + {\rm perm.}\right] ,\nonumber\\
T_{3n} & = & T_{33}  + 2 T_{34} + \frac{7}{3} T_{35}  \nonumber\\
&&+ \left[ T_{3B}\left( \ln \frac{K}{2 k_3}+\frac{7}{3}\right) + {\rm perm.}\right],
\end{eqnarray}
where we have used the identities Eq.~(\ref{eqn:Tcancel2}), (\ref{eqn:Tcancel3}) in eliminating terms and the $T_{i3}$ terms come from integrating the $x\sin x$ pieces of the $W_{i3}$ windows by parts.
  Note that the power spectrum in the
same approximation is
\begin{eqnarray}
\Delta^2_\curv(k) &\approx& e^{G(\ln s_f)} \nonumber\\
&\approx&  \frac{1}{f^2(\ln s_f)} \left[ 1+ \frac{1}{3}(n_s-1)\right],
\label{eqn:power1}
\end{eqnarray} 
where $s_f = x_f/k = e^{7/3-\gamma_E}/2k$. 
Eqn.\ (\ref{eqn:srfull})  reproduces and generalizes a well-known result for canonical ($c_s=1$, $\Xi=0$) inflation  \cite{Maldacena:2002vr, Seery:2005wm}
\begin{eqnarray}
\frac{\cal G}{k_1 k_2 k_3} 
&=& \frac{1}{k_1 k_2 k_3} \Bigg[ \epsilon_H\Big( \frac{1}{8} \sum_i k_i^3 + \frac{1}{8}\sum_{i\ne j} 
k_i k_j^2 \nonumber\\
&& + \frac{1}{K} \sum_{i>j} k_i^2 k_j^2 \Big)- \frac{\eta_H}{4} \sum_i k_i^3  \Bigg].
\end{eqnarray}
Furthermore, for squeezed bispectra 
\begin{eqnarray}
f_{\rm NL} &\equiv & \frac{5}{3} \frac{A_s^2}{\Delta_\curv^2(k_L) \Delta_\curv^2(k_S)}
 \frac{\cal G}{k_L^3} 
 \nonumber\\
 &=& \frac{5}{12}( 4\epsilon_H - 2\eta_H + \sigma_1) + {\cal O}\left( \frac{k_S}{k_L}\right)^2
\end{eqnarray}
as expected, while 
for equilateral bispectra
\begin{align}
\label{eqn:fnleq}
\frac{108}{35} f_{\rm NL}^{\rm eq} \equiv & \frac{24}{7}\left[ \frac{ A_s}{\Delta_\curv^2(k)}\right]^2 \frac{ \mathcal{G}}{k^3} \\
\approx & \frac{1}{\Delta_\curv(k) f}
\frac{c_s}{ a H s} \left( 1- \frac{1}{c_s^2}  +\frac{4}{21} \Xi \right)\Bigg|_{s=\frac{e^{-\gamma_E}}{3k}} \nonumber\\
& +\frac{2}{7} \left( \frac{c_s}{ a H s}\frac{\Xi}{f} \right)' f + \frac{7c_s^2+8}{7 c_s^2} \epsilon_H + \frac{7 c_s^2-10} {7 c_s^2} {\sigma_1} \nonumber\\
& + \frac{2(c_s^2-10) - 135(c_s^2-1) \ln(3/2)}{14c_s^2}(n_s-1). \nonumber
\end{align}

%
%

The latter should be compared with
the zeroth order in slow-roll result
\begin{equation}
\frac{108}{35} f_{\rm NL}^{\rm eq0} =\left( 1- \frac{1}{c_s^2} +\frac{4}{21} \Xi\right) \Bigg|_{s=\frac{1}{3 k}}.
\label{eqn:fnleq0}
\end{equation}
For example, for $\Xi=0$, we obtain using  Eq.~(\ref{eqn:power1})
\begin{align}
\frac{f_{\rm NL}^{\rm eq}- f_{\rm NL}^{\rm eq0}}{f_{\rm NL}^{\rm eq0}} 
=&\frac{8 c_s^2 -17 - 64(c_s^2-1) \ln(3/2)}{7 (c_s^2-1)}(n_s-1) \nonumber\\
&+\frac{15 \epsilon_H  + (14\gamma_E-3)\sigma_1 }{7 (c_s^2-1)} .
\label{eqn:fnleqcorrected}
\end{align}
Note that as one would expect, at small $c_s^2$ the correction is of order the deviation from scale
invariance, $n_s-1$, unless $\epsilon_H$ and $\sigma_1$ are anomalously larger.  
As $c_s^2 \rightarrow 1$, the zeroth order term is itself suppressed and 
$f_{\rm NL}^{\rm eq0}$ drops below the slow-roll corrections.   Here the correction is
large but the equilateral bispectrum itself is small.

\begin{figure}[t]
\psfig{file=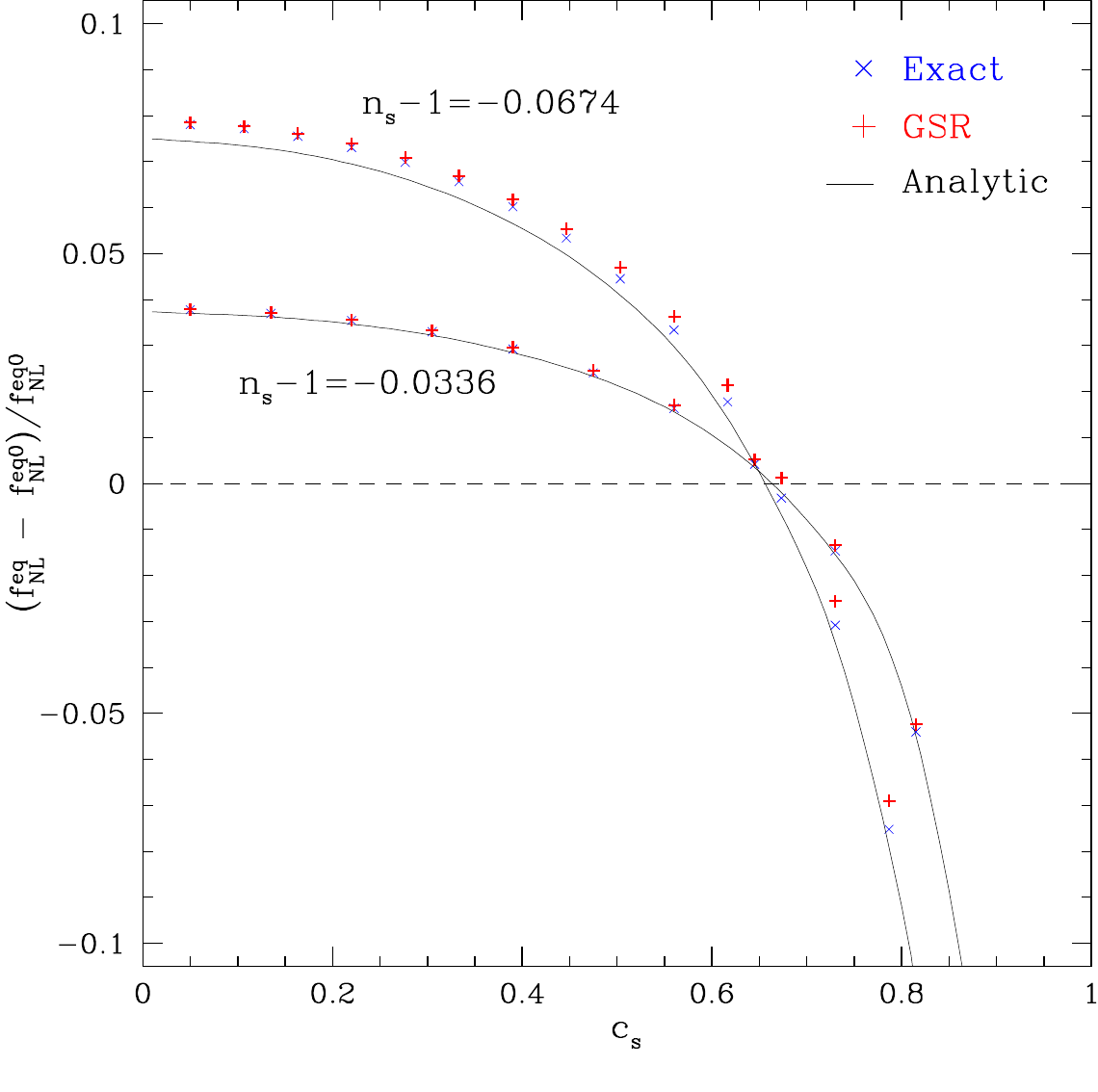, width=3in}
\caption{Slow roll corrections to the zeroth order equilateral bispectrum result
$f_{\rm NL}^{\rm eq0}$.   Agreement between the exact, GSR and analytic 
results is everywhere excellent for realistic values of the tilt $n_s-1= -0.0337$.
Shown also for comparison is a case of twice the observed value $n_s-1=-0.0674$
showing that corrections are always of order the tilt.   They only become
fractionally large as $c_s \rightarrow 1$ as the zeroth order result becomes suppressed.
}
\label{fig:sr}
\end{figure}

\subsection{Examples}\label{app:litcomp}

We again illustrate our calculations with the DBI example, here with no feature $(b=0)$.
In Fig.~\ref{fig:sr}, we  show  the exact, GSR and first order corrected analytic result of
Eq.~(\ref{eqn:fnleqcorrected})  relative
to the zeroth order $f_{\rm NL}^{\rm eq}$ of Eq.~(\ref{eqn:fnleq0}).  Compared with 
the zeroth order result, the corrections are of order the tilt $n_s-1$ throughout and are
only fractionally large for $c_s >0.8$ where the zeroth order term itself drops below
$0.2$.  The agreement between the exact, GSR and first order results are is excellent
and consistent with deviations only at the next order in the slow-roll expansion
$(n_s-1)^2$.   By taking a second artificially high $n_s-1$, we also demonstrate that
first order corrections are always of order the tilt in these cases where the individual slow-roll
parameters are not anomalously larger than tilt.   For both cases, most of the $(n_s-1)^2$
deviation  between the analytic and exact or GSR results can in fact be corrected by
evaluating Eq.~(\ref{eqn:fnleq}) at the exact freezeout epoch specified rather than using
the approximate conversion in Eq.~(\ref{eqn:fnleqcorrected}).

We can also  compare the results derived here to known slow-roll results in the literature. In particular, we will compare to Ref.~\cite{Burrage:2011hd}. Since the expressions are rather large and cumbersome we will restrict again to the equilateral limit of DBI. Furthermore, the results of Ref.~\cite{Burrage:2011hd} are evaluated as an expansion about a fixed reference point. That is, while the dependence on wavenumber of coefficients in our Eq.\ (\ref{eqn:fnleq}) is implicit, their coefficients are constant so that the resulting dependence on wavenumber is explicit.  

To compare results, we need to expand our expressions about some reference scale  $s_{\star}$. Quantities evaluated at this time will be denoted with a subscript $\star$.  Working to linear order in slow-roll parameters and making use of Eq.\ (\ref{eqn:srcsrexp}), we find
\begin{align}\nn
 \left.\frac{1}{f} \frac{c_s}{ a H s}\(\frac{1}{c_s^2} - 1\)\right|_{x_f}
 \approx &  \(\frac{1}{c_{s\star}^2} - 1\)\frac{1}{{f_\star}} \Big[
 {1-\epsilon_{H\star }- \sigma_{1\star}}\\ & -\(\frac{1-n_{s\star}}{2} + \frac{2\sigma_{1\star}}{ 1-c_{s\star}^2}\) 
 \nn\\
&\quad\times \ln\(3 k s_\star e^{\gamma_E}\)\Big],
\end{align}
where $x_f  =  3k s_f  = e^{-\gamma_E}$. Finally, we make use of the result,
\begin{align}
\frac{1}{{\Delta_{\curv}(k)}}\frac{1}{f_{\star}}
=  1-\frac{2  - \ln\(2 k s_{\star}e^{\gamma_E}\)}{2}(1-n_{s\star})
\end{align}
from Eq.~(\ref{eqn:power1}). Combining terms, for DBI inflation we find
\begin{align}\label{eqn:fnlDBISR}
f_{\rm NL}^{\rm eq}
=& \frac{5}{108 \csss} \Big\{  7 (\csss - 1) + 
15 \epsilon_{H\star} \nonumber\\
& +\left[ 8\csss-17-64(\csss-1)\ln(3/2) \right](n_{s\star}-1) \nonumber\\
& +\left[ 14\ln(3 k s_\star e^{\gamma_E} )-3\right] \sigma_{1\star} \Big\}.
\end{align}
In the limit where DBI inflation is described by power-law inflation at late times \cite{Silverstein:2003hf}, $\sigma_1 = -2 \epsilon_H$, $\eta_H = \epsilon_H$, and Eq.\ (\ref{eqn:fnlDBISR}) reduces to
\begin{align}
f_{\rm NL}^{\rm eq}  = \frac{35}{108}\[\(1-\frac{1}{c_{s\star}^2} \)+\frac{3-4\ln(3ks_{\star}e^{\gamma_E})}{c_{s\star}^{2}}\epsilon_{H\star}\],
\end{align}
in agreement with Eq.\ (6.5) of Ref.~\cite{Burrage:2011hd}. We can also check the example of so-called generalized DBI inflation \cite{Franche:2009gk}. The slow-roll parameters
there are defined as
\begin{align}
\epsilon_V = \frac{1}{2}\(\frac{V'}{V}\)^2, \; \eta_{V} = \frac{V''}{V}, \; \Delta = \frac{{\rm sgn}(\dot\phi T^{1/2})}{3H}\frac{T'}{T^{3/2}},
\end{align}
and are related to our slow-roll parameters by \cite{Burrage:2011hd}
\begin{align}\nn
\epsilon_H = c_{s}\epsilon_V, \;  \eta_{H} =  &-\frac{c_{s}}{2}\epsilon_V + \frac{c_s}{2}\eta_{V}+\frac{3}{4}\Delta\\
\sigma_{1} = -c_{s}\epsilon_V&+c_{s}\eta_V - \frac{3}{2}\Delta.
\end{align}
Note that their $\eta = 2(\epsilon_{H} - \eta_H)$.
Working in the limit $c_{s}\ll 1$, we find
\begin{align}\nn
f_{\rm NL}^{\rm eq}= & -\frac{35}{108c_{s\star}^2}\Bigg[1+ \frac{\eta_{V\star}c_{s\star}}{7}\[3- 14\ln\(3k s_\star e^{\gamma_E}\)\] \\ \nn& +\frac{3\Delta_\star}{14}\[31+14 \ln\(3 k s_{\star} e^{\gamma_E}\)- 128 \ln (3/2)\]\\\nn & 
 -\frac{2\epsilon_{V\star}c_{s\star}}{7}\[43  - 7\ln\(3k s_\star e^{\gamma_E} \)- 128 \ln(3/2)\] 
\Bigg]\\
& + \mathcal{O}(c_{s\star}^{2})
\end{align}
in agreement with Eq.\ (6.11) of Ref.~\cite{Burrage:2011hd}.

\section{DBI Parameters}
\label{app:parameters}

In the main text, we choose the DBI parameters to have a fixed power spectrum
amplitude $A_s$ and tilt $n_s-1$ at $s_s$ for a model with no step $b=0$.  
 This condition fixes the parameters
 $\{\phi_{\rm end},\lambda_B,\beta\}$ of the model and we set $V_0$ to a constant 
 as described in the main text.   We relate the phenomenological
 and fundamental parameters through the slow-roll attractor
\begin{equation}
 H \frac{d\phi}{d \ln a}= \sqrt{2 X} \approx -\sqrt{ \frac{V}{3}} \frac{d\ln V}{d\phi} c_s ,
\label{eqn:attractor}
\end{equation}
which determines $X$ at $\phi$,
and the definition of the sound speed $c_s = \sqrt{1-2 X/T}$.
Through this relation, the phenomenological parameters $\bp = \{A_s,n_s-1 ,c_s\}$ are defined
given the field position which completes the set of 
fundamental parameters
$\bpi=\{ \phi,\lambda_B,\beta\}$.  Explicitly, we set
\begin{eqnarray}
c_s(\bpi) &=& \sqrt{1-2 X/T} \nonumber\\
&\approx &
\left[ 1+ \frac{V}{3 T}\left( \frac{ d\ln V}{d\phi}\right)^2  \right]^{-1/2}, 
\end{eqnarray}
and use the Friedmann equation to set $H$
\begin{equation}
3H^2(\bpi) =  \left(  \frac{1}{c_s} -1 \right)T(\phi)  + V(\phi) \,.
\end{equation}
The amplitude and tilt are then given as
\begin{eqnarray}
A_s(\bpi) &\approx& \left(\frac{H}{2\pi d\phi/d\ln a}\right)^2 ,\nonumber\\
(n_s-1)(\bpi) &\approx&  \frac{d\ln A_s}{d \phi}\frac{d\phi}{d\ln a}.
\end{eqnarray}
This gives the phenomenological parameters as a nonlinear function of 
the fundamental parameters $\bp(\bpi)$.
To set the sound horizon at $\phi$ to equal $s_s$, we choose the appropriate
$\phi_{\rm end}$ such that we effectively replace $\phi$ with $\phi_{\rm end}$ in 
the fundamental parameter set after the fact.   

Of course, we actually want the fundamental parameters as a function of the 
phenomenological parameters $\bpi(\bp)$.    For $c_s\ll1$, these relations are easily
inverted
\begin{align}
\lambda_B \approx& \frac{1}{4\pi^2 A_s} \left( \frac{4}{n_s-1} \right)^4 ,\nonumber\\
\beta \approx & -\frac{3}{4}\frac{n_s-1}{c_s} ,\nonumber\\
\phi \approx & \frac{2\sqrt{3}}{3\pi} \sqrt{\frac{V_0}{A_s (n_s-1)^2}}.
\end{align}
For larger $c_s$, the expressions are not readily invertible but from the solution at small
$c_s$, we can approximate small changes by linearizing and inverting the
response.  Starting from some parameter set $\bp_0$, we move to a new set $\bp$
by iterating 
\begin{equation}
\ln(\bpi/\bpi_0) = \left( \frac{ \partial\ln \bp }{\partial\ln \bpi}\right)^{-1} \ln(\bp/\bp_0),
\end{equation}
until convergence.  Here the inverse factor is the Jacobian matrix inverse.  
We repeat this procedure until we obtain all  the desired values of the sound speed $c_s$.

\vfill \break
\bibliography{DBIBi}

\end{document}